\documentclass[11pt,journal,draftclsnofoot,letterpaper,onecolumn]{IEEEtran}
\usepackage{endnotes}
\usepackage{graphicx}
\usepackage{dcolumn}
\usepackage{bm}
\usepackage{amsmath,amsfonts, amssymb}
\usepackage{latexsym}
\usepackage{lineno}
\usepackage{array}
\usepackage{adjustbox}
\usepackage[normalem]{ulem}
\usepackage{xcolor, soul}
\usepackage[hidelinks]{hyperref}
\usepackage{booktabs}
\usepackage{caption}
\usepackage{subcaption}
\usepackage{float}
\usepackage{algorithm} 
\usepackage[noend]{algpseudocode} 
\usepackage{multirow}
\usepackage{url,hyperref,lineno,microtype,subcaption}
\usepackage[onehalfspacing]{setspace}

\def\Authors{Talmon Alexandri\,$^{1,\href{https://orcid.org/0000-0003-3778-9298}{\includegraphics[scale=0.05]{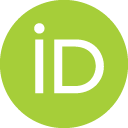}}}$, Roee Diamant\,$^{1~\href{https://orcid.org/0000-0002-2430-7061}{\includegraphics[scale=0.05]{ORCID128.png}}}$}

\DeclareMathOperator*{\argmax}{argmax}

\def\ve#1{{\mathchoice{\mbox{\boldmath$\displaystyle #1$}}%
		{\mbox{\boldmath$\textstyle #1$}}%
		{\mbox{\boldmath$\scriptstyle #1$}}%
		{\mbox{\boldmath$\scriptscriptstyle #1$}}}}
\usepackage{etoolbox}
\usepackage{tikz}
\newlength{\radius}
\newrobustcmd*{\mytriangle}[1]{\tikz{\draw[thin, fill=#1] (90:\radius) -- (210:\radius) -- (330:\radius) -- cycle;}}
\newrobustcmd*{\myuptriangle}[1]{\tikz{\draw[thin, fill=#1] (270:\radius) -- (30:\radius) -- (150:\radius) -- cycle;}}

\makeatletter
\newcommand{\new}[1]{{\textcolor{black}{#1}}}
\makeatletter
\newcommand{\newnew}[1]{{\textcolor{black}{#1}}}

\makeatletter
\newcommand\remembertext[2]{
	\immediate\write\@auxout{\unexpanded{\global\long\@namedef{mytext@#1}{#2}}}%
	#2
}
\newcommand\recalltext[1]{%
	{\ifcsname mytext@#1\endcsname
		\@nameuse{mytext@#1}%
		\else
		``??''
		\fi
}}
\makeatother

\begin{document}

\onecolumn

\title{Design of an Optimal Testbed for Tracking of Tagged Marine Megafauna} 
\author{\Authors\thanks{\indent Talmon Alexandri (email: talexa03@campus.haifa.ac.il) and Roee Diamant (email: roee.d@univ.haifa.ac.il) are with the Department of Marine Technologies, University of Haifa, Haifa, Israel }} 


\maketitle

\section{Abstract}
Underwater acoustic technologies are a key component for exploring the behavior of marine megafauna such as sea turtles, sharks, and seals. The animals are marked with acoustic devices (tags) that periodically emit signals encoding the device's ID \remembertext{r4c2}{\newnew{along with sensor data such as}} depth, temperature, or the dominant acceleration axis - data that is collected by a network of deployed receivers. \remembertext{r1c1}{\new{In this work, we aim to optimize the locations of receivers for best tracking of acoustically tagged marine megafauna.}}The outcomes of such tracking allows the evaluation of the animals' motion patterns, their hours of activity, and their social interactions. In particular, we focus on how to determine the receivers' deployment positions to maximize the coverage area in which the tagged animals can be tracked. For example, an overly-condensed deployment may not allow accurate tracking, whereas a sparse one, may lead to a small coverage area due to too few detections. We formalize the question of where to best deploy the receivers as a non-convex constraint optimization problem that takes into account the local environment and the specifications of the tags, and offer a sub-optimal, low-complexity solution that can be applied to large testbeds. Numerical investigation \remembertext{r1c0}{\new{for three stimulated sea environments}}shows that our proposed method is able to increase the localization coverage area by 30\%, and results from a \remembertext{r1c2}{\new{test case}}experiment demonstrate similar performance in a real sea environment. We share the implementation of our work to help researchers set up their own acoustic observatory.

\section{Introduction}\label{sec:intro}
Underwater acoustic tracking is a key enabling technology for exploring long-term behavior of marine megafauna \cite{kraus2018evaluation}. \remembertext{r1c3}{\new{Acoustic telemetry of fish developed in the mid 1950s by the U.S. Bureau of Commercial Fisheries (BCF), enabling the identification and localization of the individual fish without the need to recapture it \cite{hockersmith2012history}. Tracking of decapod crustaceans with acoustic telemetry devices evolved since the 1970s and as of toady 60\%  of published studies are based on acoustic telemetry \cite{florko2021tracking}, mostly using acoustic tags. \newnew{Although tracking of megafauna can be performed by passive and active system, e.g., \cite{diamant2019active}, the data obtained from acoustic tagging is far more informative. As a result, acoustic tagging is used in multitude of} research projects such as the ocean tracking network (OTN) (https://oceantrackingnetwork.org/). For marine megafauna, acoustic tags have been used for understanding the behavioral and social interactions of animals like sharks, sea turtles, and seals \cite{lea2016acoustic,berejikian2016predator}.}}Anchored receivers are deployed in known locations within the explored area to decode and measure the time of arrival (ToA) of the tags' emissions for, usually offline, tracking of the tagged animals. The long-term durability of the tags and receivers that can operate for many months and years allows operation over a long period of time for the statistical evaluation of the activity of the tagged animals. Examples of tagged sharks \cite{lea2016acoustic} and turtles \cite{thums2013tracking} revealed valuable information about the animals' local motion patterns, their hours of activity, and their social interactions. 

\remembertext{r1c4}{\new{The detection of the animal's tag emissions using a single receiver is sufficient for collecting indications of presence/absence of an animal in an area of interest \cite{gazit2013deployment, jackson2011development}. However, in the case of fixed receivers, receptions of a tag's emissions by at least three receivers is required to localize the animal in a two-dimensional (2D) plane with no ambiguities.}}Such localization involves time synchronizing the receivers and fusing their ToA data collection in a time difference of arrival (TDoA) localization framework \cite{alexandri2018tracking}. In this context, a key for obtaining a large coverage area for tracking the tags is the positioning of the anchored receivers. \remembertext{r1c5}{\new{Specifically, on one hand, overly-close deployment would decrease the localization accuracy due to small angular difference between the tag to be localized and each of the receivers. On the other hand, an overly-wide deployment would yield a small coverage area due to the limited transmission range of the emitting tag.}}This problem is further exacerbated when \remembertext{r1c6}{\new{the bathymetry of the explored area is complex and the propagation loss for the tag-receiver link depends on the varying water column temperature and depth change of the seabed}}. As a result, not only the relative distance between the receiver should be considered, but also the receivers' geographic position \cite{huveneers2016influence}. 

When designing a testbed for the acoustic tracking of tagged marine megafauna, the anchors are commonly spaced according to the declared detection range \cite{as2021pinpint}, and are spread geographically to best cover an \textit{area of interest}, $\mathcal{I}$. Deployment strategies include a line of receivers to detect tagged individuals passage through a river or along a coastline \cite{kraus2018evaluation}, or an array of evenly-spread receivers \cite{as2021pinpint, kessel2014review}. A common practice (CP) for positioning the receivers is to cover the area of interest using equilateral triangles \cite{shiu2010divide}, whose edges are set to half of the detection range. This practice stems from the fact that, under a simplified scenario of a cylindrical propagation loss, a close-to-optimal solution is obtained (see analysis in the Appendix). However, in the practical case where the seabed is complex and the detection range is not iso-symmetric, a more rigorous way to determine the anchors' deployment location is required. Two examples of the under-utilization of the deployment setup using CP are a testbed setup to track acoustically tagged slipper lobsters, \textit{Scyllarides latus}, at the Achziv Marine Nature Reserve in northern Israel \cite{alexandri2018tracking}, and a testbed aimed to explore the motion of tagged sandbar sharks, \textit{Carcharhinus plumbeus}, close to the ``Orot Rabin'' Power and Desalination Station in Hadera, Israel $(32^028'N; 34^052'E)$~ \cite{alexandri2019localization}. In the former, four receivers were anchored in a triangle according to the CP approach over a rocky seabed, and 19 lobsters were tagged and tracked for a period of 8~months. Out of more than 45,000 detected tags' emissions, only 252 \remembertext{r1c7a}{\new{(0.5\%)}}were received by 3 or 4 receivers. In a second testbed, 20 tagged sharks were tracked for a period of 87 days by a set of 4 receivers deployed on a shallow sandy seabed according to CP. Out of a total of 42,589 detected tags' emissions, only 180 \remembertext{r1c7b}{\new{(0.4\%)}}emissions were received by 3 receivers to enable localization. \remembertext{r1c7}{\new{A summary of the poor results due to under-utilization of CP deployment setup is presented Table \ref{tab:receptions}.}}

\begin{table}[ht]
	\caption{\new{Records of successful tag detections from Achziv Marine Nature Reserve tracking 19 slipper lobsters for a period of 8 month, and from “Orot Rabin” Power and Desalination Station testbed tracking 20 sandbar sharks for a period of 87~days. Only 0.4\% to 0.5\% of received tags' emissions were received by 3 or more receivers. Results demonstrate the under-utilization of CP deployment setup.}}
	\new{
		\begin{center}
			\begin{tabular}{lcc}
				\toprule
				Testbed  & Achziv & “Orot Rabin”\tabularnewline
				\midrule
				Number of received emissions by single receiver  & 45,000 & 42,589 \tabularnewline
				\midrule
				Number of received emissions by three or more receivers&  252 &  180 \tabularnewline
				\midrule
				Percentage of received emissions  by three or more receivers [\%] & 0.5 & 0.4 \tabularnewline
				\bottomrule
			\end{tabular} 
			\label{tab:receptions}
		\end{center}
	}
\end{table}	  

The problem of how to position receiving nodes resembles the "Art gallery problem" \cite{rajagopal2016beacon}, whose goal is to find the minimal set of guards such that every point in a floor plan is covered by at least one guard. In the case of 2D localization, three receivers represent a guard. While this type of optimization problem is proven to be NP-Hard (nondeterministic polynomial time hard) \cite{lee1986computational}, \remembertext{r1c8a}{\new{i.e., the complexity of search increases with the problem size in such a way that a solution would compare with the brute-force approach of trying out the entire state-space,}}some work-arounds are possible. For indoor localization, information about the floor plan is utilized to reconcile position ambiguities formed by the localization of a node by only two receivers \cite{rajagopal2016beacon}. The proposed method disqualifies potential positions located outside the floor or behind barriers/walls that block the signals. To manage spatial-dependent propagation loss in indoor localization, templates of the reception pattern are used \cite{mnasri2015genetic} as well as modeled calculations \cite{rajagopal2016beacon}. However, the solution for underwater localization is somewhat different. First, while the sea includes barriers that can be used to resolve localization ambiguities, these mostly apply in near shore locations. Furthermore, while some solutions relied on the structure of the seabed for localization, e.g., in \cite{dubrovinskaya2017anchorless}, localization accuracy is low if the structure is not diverse.

In this paper, we propose a systematic method to determine the deployment setup of a marine megafauna tracking testbed considering the acoustic specifications of the tags, the number of receivers, and the environmental properties. Relying on prior knowledge of the bathymetric and bathythermal conditions in the explored area, we formalize a constraint optimization problem that accounts for the spatially dependent propagation loss, and yields the best deployment locations to maximize the coverage area. In our analysis, we consider the user's area-of-interest, $\mathcal{I}$, and evaluate the coverage quality by the geometrical dilution of precision (GDOP) \cite{sharp2009gdop}. Since $\mathcal{I}$ may be non-convex, the optimal solution is often NP-hard, \remembertext{r1c8b}{\new{ i.e. to solve the decision problem of "where to position the receivers?" requires unreasonable long time such that}}even for a modest $\mathcal{I}$, the solution becomes too hard to evaluate. For these cases, we offer a sub-optimal implementation based on a genetic algorithm (GA). We refer to our approach as the \textit{Propagation-Dependent Anchor Deployment} (PDAD) scheme. For simple setups, we show that PDAD achieves the optimal solution, as verified by a brute force search. For more complex setups, we compare the coverage area yielded by PDAD to that of CP.

Our contribution is twofold:
\begin{enumerate}
	\item A novel systematic approach and optimization formalization for how to set up the location of anchored receivers in an underwater testbed.
	\item A method to merge the GDOP metric with bathymetric information to quantify detection ranges.
\end{enumerate}

We explore the performance of our method in numerical simulations and in a sea experiment. The former demonstrates the obtained deployment strategy for three different seabed environments and explores the sensitivity of the results to the system's parameters, as well as the degree of sub-optimality of the proposed solution. The latter demonstrates the merit of using the proposed approach in a real sea environment. Compared to the CP strategy, the results show an increase of roughly 30\% in the size of the covered area as obtained by PDAD. 

\section{Materials and Methods}

In this section, we describe our system model and formulate our solution. \remembertext{r4c7}{\newnew{We use the following notations: coordinates’ are marked by a macron sign, e.g., $\bar{p}$, sets are marked by bold letters, e.g., $\ve{D}_\mathrm{i}$.}} Table \ref{tab:abb} summarizes the paper's main variables and notations.

\subsection{Study Area}

Our setup includes $N$ receivers deployed to cover a given area-of-interest, $\mathcal{I}$. Our goal is to determine the optimal location and deployment depth for the receivers, such that a tagged animal that passes through $\mathcal{I}$ is well-localized within area $\mathcal{I}$ and possibly beyond it. For each tag's emission, the detecting receiver measures the local arrival time, such that, assuming all $N$ receivers are time-synchronized, localization is performed by TDoA cf. \cite{alexandri2018tracking}. 

We assume prior knowledge to evaluate the propagation loss between any pair of positions within and beyond $\mathcal{I}$. In particular, we require information about the bathymetry within the explored area and the expected bathythermal profile of the water. This information is used to run a propagation loss model such as ray-tracing, normal modes, or parabolic equation \cite{bahrami2016underwater,gul2017underwater}. We admit that, since the bathythermal profile is time varying and depends on seasons, the sound velocity profile (SVP) cannot be accurately known for the entire deployment time period. Yet, a track of the bathythermal profile may reveal temporal trends in the sound speed, allowing the evaluation of a \textit{nominal SVP}. Alternatively, the user may perform periodic measurements of the bathythermal and change the locations of the receivers if needed.

\begin{table}[H]
	\caption{Main variables and notations - Coordinates' variables are marked by a macron sign, e.g., $\bar{p}$. Sets are marked by bold letters, e.g., $\ve{D}_\mathrm{i}$.  }
	\begin{center}
		\begin{tabular}{>{\centering\arraybackslash}m{2cm}>{\arraybackslash}m{12cm}}
			\toprule 
			Variables & Descriptions\tabularnewline
			\midrule
			$\alpha$ & User-defined GDOP localization quality threshold \tabularnewline 
			$\beta$ & Number of required receivers for localization \tabularnewline
			i & Index of a receiver and its associated detection area\tabularnewline
			j & Receivers set index whose detection areas intersect \tabularnewline
			$\bar{r}_\mathrm{i}$ & Position of receiver number i \tabularnewline
			$\bar{p}$            & Position of an emitting source \tabularnewline
			$\ve{D}_\mathrm{i}$  & Detection area of the $\mathrm{i}^{th}$ receiver as defined by \eqref{eq:pl} \tabularnewline
			$\ve{\lambda}_\mathrm{j}$ & Set of receivers whose detection areas intersect as per \eqref{eq:lamb} \tabularnewline
			$\ve{\Lambda}$ & A group containing sets $\ve{\lambda}_\mathrm{j}$ \tabularnewline
			$\ve{L}_\mathrm{j}$  & Localization area resulting from the intersection of $\beta$ detection areas. Defined in \eqref{eq:loc} \tabularnewline
			$\ve{\mathcal{U}}_\mathrm{j}$  & Set of all positions of the source, $\bar{p}$, in the localization
			area, $\ve{L}_\mathrm{j}$, in which $\mathrm{GDOP}(\bar{p}, \ve{L}_\mathrm{j}) \leqslant \alpha$. Defined in \eqref{eq:use} \tabularnewline
			$\ve{\mathcal{C}}$ & Coverage area, a union of all J usable areas \tabularnewline
			$\ve{\mathcal{I}}$ & Area-of-interest, area to be covered by the optimized deployment \tabularnewline
			GDOP & Geometrical delusion of precision  \tabularnewline
			$\eta$ & Ratio between the area covered by deployment according to PDAD to deployment according to CP \tabularnewline
			$\theta$ &  Ratio between the coverage and the usable areas \tabularnewline[0.1cm]
			$P_{\mathrm{SL}},~ P_{\mathrm{TL}}$ & Source level power and transmission loss respectively \tabularnewline
			$P_{\mathrm{NL}},~ P_{\mathrm{SNR}}$ & Noise level power and received signal to noise respectively \tabularnewline
			$\xi_{i, r}$ & Throughput of the tags - receivers link \tabularnewline
			\bottomrule
		\end{tabular}
	\end{center}
	\label{tab:abb}
\end{table}
\clearpage
\subsection{Detection Range}
Let $P_\mathrm{SL}$ be the source power level of the tag to-be-localized. Let $P_\mathrm{NL}$ be the ambient noise level, assumed constant in the explored area, and let $P_{\mathrm{TL}}(\bar{r}_i,\bar{p})$ be the power transmission loss between a source node at position $\bar{p}=(x,y,z)$ and the $i$th receiver at position $\bar{r}_\mathrm{i} = (x_\mathrm{i}, y_\mathrm{i}, z_\mathrm{i}), ~ \mathrm{i = 1, 2, ..., N}$. We calculate $P_\mathrm{TL}(\bar{r}_\mathrm{i},\bar{p})$ and $P_{\mathrm{NL}}$ by acoustic channel modeling, \remembertext{r1c12a}{\new{i.e., modeling the acoustic attenuation by considering the environmental conditions, the ambient noise, and the emitted frequency,}}e.g., \cite{gul2017underwater}, and from the Wenz curves for acoustic ambient noise in the ocean \cite{national2003ocean}, \remembertext{r1c12b}{\new{showing the average ambient noise spectra for different levels of shipping traffic, and sea state conditions. Specifically, analyzing a huge set of acoustic measurements of ambient noise, Wenz was able to provide an empirical curve for the ambient noise level and to show that it is frequency and environment dependent.}}Then, comparing the received level
\remembertext{r1c13}{\new{
\begin{eqnarray}
	P_\mathrm{SNR} = \frac{P_\mathrm{SL}\times P_\mathrm{TL}(\bar{r}_\mathrm{i},\bar{p})}{P_\mathrm{NL}}
\end{eqnarray}
}}

\noindent to a detection threshold, $P_\mathrm{DT}$, we measure whether location $\bar{r}_\mathrm{i}$ is suitable for detecting a source located at $\bar{p}$. \remembertext{r1c14}{\new{That is, if $P_\mathrm{SNR} \geqslant P_\mathrm{DT}$ a receiver positioned at $\bar{r}_\mathrm{i}$ will detect a tag emitting at $\bar{p}$.}}

\subsection{Localization Quality}
A common metric to measure the achievable localization accuracy for the planned receivers' deployment is the GDOP. This is a unitless metric that ranks the deployment setup by considering both the measurement precision and the geometry between the source and receivers to account for the effect of the geometry setup on localization \cite{kaplan2005performance}. \remembertext{r1c15}{\new{As localization quality improves, the GDOP decreases. Classification of GDOP values are outlined in table Table \ref{tab:GDOP}.}}For example, if all receivers are colinear, the achievable localization quality is poor. Given a set $\ve{\lambda} = \{\bar{r}_1, \bar{r}_2, ..., \bar{r}_N\}$ of N receivers stationed at locations $\bar{r}_1, \bar{r}_2, ..., \bar{r}_N$, the GDOP is defined as the ratio between the accuracy of a position fix to the variance of the measurements \cite{sharp2009gdop}. Formally, denote the visibility matrix \cite{yarlagadda2000gps, kaplan2005performance}

\begin{eqnarray}
	\ve{H}(\bar{p}, \ve{\lambda})=\left[\begin{array}{cccc}
		a_\mathrm{x1} & a_\mathrm{y1} & a_\mathrm{z1} & 1\\
		a_\mathrm{x2} & a_\mathrm{y2} & a_\mathrm{z2} & 1\\
		a_\mathrm{x3} & a_\mathrm{y3} & a_\mathrm{z3} & 1\\
		\vdots        &\vdots         &\vdots         &\vdots\\
		a_\mathrm{xN} & a_\mathrm{yN} & a_\mathrm{zN} & 1
	\end{array}\right]
\end{eqnarray}
whose elements are the unit vectors pointing from a potential position of the source, $\bar{p}=[p_\mathrm{x},~p_\mathrm{y},~p_\mathrm{z}]$, to the location of the $\mathrm{i^{th}}$ receiver, $\bar{r}_\mathrm{i}=[r_\mathrm{x_i},~r_\mathrm{y_i},~r_\mathrm{z_i}]$, such that $a_\mathrm{xi} = (p_\mathrm{x}-r_\mathrm{x_i})/R_\mathrm{i}$, $a_\mathrm{yi} = (p_\mathrm{y}-r_\mathrm{y_i})/R_\mathrm{i}$, and $a_\mathrm{zi} = (p_\mathrm{z}-r_\mathrm{z_i})/R_\mathrm{i}$, where $R_\mathrm{i}=\sqrt{(p_\mathrm{x}-r_\mathrm{x_i})^2+(p_\mathrm{y}-r_\mathrm{y_i})^2+(p_\mathrm{z}-r_\mathrm{z_i})^2}$, all in Cartesian coordinates. The $\mathrm{GDOP}(\bar{p}, \ve{\lambda})$ equals $\sqrt{g_\mathrm{11}^\mathrm{2}+g_\mathrm{22}^\mathrm{2}+g_\mathrm{33}^\mathrm{2}+g_\mathrm{44}^\mathrm{2}}$, where
\begin{eqnarray}
	\left(\ve{H}^\mathrm{T}\ve{H}\right)^\mathrm{-1}=\left[\begin{array}{cccc}
		g_\mathrm{11} & g_\mathrm{12} & g_\mathrm{13} & g_\mathrm{14}\\
		g_\mathrm{21} & g_\mathrm{22} & g_\mathrm{23} & g_\mathrm{24}\\
		g_\mathrm{31} & g_\mathrm{32} & g_\mathrm{33} & g_\mathrm{34}\\
		g_\mathrm{11} & g_\mathrm{22} & g_\mathrm{33} & g_\mathrm{44}
	\end{array}\right]\;.
\end{eqnarray}

\subsection{Problem Statement}
\begin{figure}[h]
	\centering
	\includegraphics[width=0.95\columnwidth]{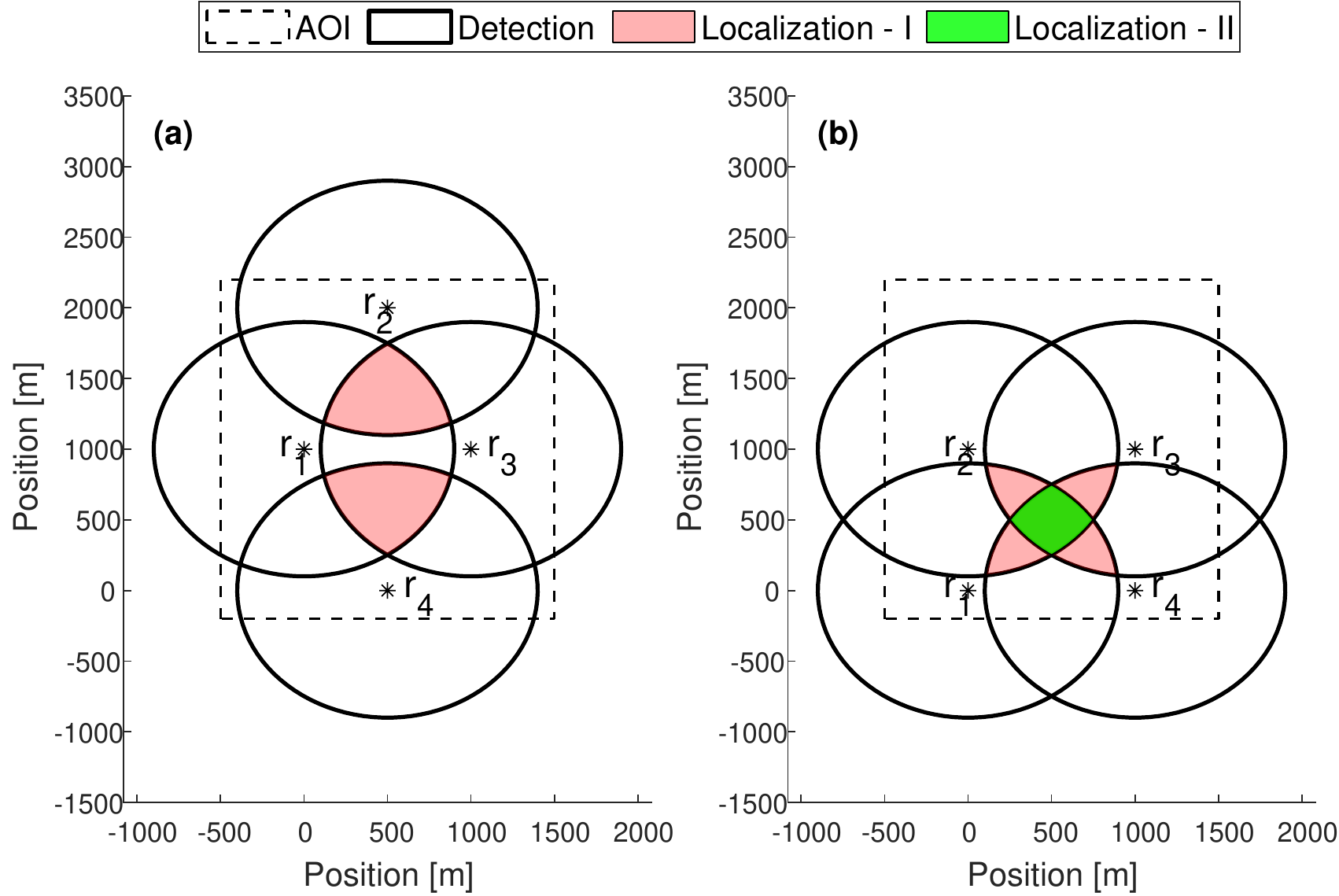}
	\caption{An example showing the deployment of $N=4$ receivers $\bar{r}_1,~\bar{r}_2,~\bar{r}_3,~\bar{r}_4$ with corresponding detection areas marked by black contour lines. For the case of $\beta = 3$, in deployment setup (a), although $J=5$, there are two localization areas, $\ve{L(\lambda_1)},~\ve{L(\lambda_2)}$, marked by the light red shading. The two sets of at least 3 receivers are $\ve{\lambda_1}=\{\bar{r}_1,~\bar{r}_2,~\bar{r}_3\}$ and $\ve{\lambda_2}=\{\bar{r}_1,~\bar{r}_3,~\bar{r}_4\}$. The coverage area, $\mathcal{C}$, is the union of $\ve{L(\lambda_1)},~\ve{L(\lambda_2)}$. In the deployment setup (b), the light red areas form "Localization area - I", which merges localization areas $\ve{L(\lambda_1)},~\ve{L(\lambda_2)},~\ve{L(\lambda_3)},~\ve{L(\lambda_4)}$, for which $\beta=3$ detection areas overlap. The light green areas form "Localization area - II", for which $\beta=4$ detection areas overlap. For both deployment setups, the resulting localization area is non-convex.}
	\label{fig:zones}
\end{figure}

To formulate the receivers' deployment problem, \remembertext{r1c18}{\new{four sets of variables are required.}} Recall that the source node is detected when $P_\mathrm{DT}(\bar{r}_\mathrm{i},\bar{p}) \leqslant P_\mathrm{SL}\times P_\mathrm{TL}/ P_\mathrm{NL}$. We define the \textit{detection area}, $\ve{D}_\mathrm{i}$, as the set of all possible positions of the source such that a receiver located at $\bar{r}_\mathrm{i}$ will detect the source's transmissions. Formally, 
\begin{eqnarray}
	\bar{p}\in \ve{D}_\mathrm{i}~ \vert~ P_\mathrm{DT}(\bar{r}_\mathrm{i},\bar{p}) \leqslant \frac{P_\mathrm{SL}\times P_\mathrm{TL}(\bar{r}_\mathrm{i},\bar{p})}{P_\mathrm{NL}} \;.
	\label{eq:pl}
\end{eqnarray}  
The resolution of set $\ve{D}_\mathrm{i}$ is determined by the resolution of the bathymetric information. An example of a detection area for four receivers is shown in Fig.~\ref{fig:zones}(A). Note that the detection area obtained is not necessarily convex. Let $\ve{\Lambda}(\beta)$ be a group containing sets $\ve{\lambda}_\mathrm{j}$, j=1,\ldots,J ,where $\ve{\lambda}_\mathrm{j}$ is \remembertext{r1c19}{\new{the $j$th set}} of at least $\beta$ receivers whose detection areas intersect such that
\begin{eqnarray}
	\ve{\lambda}_\mathrm{j}\in\ve{\Lambda}(\beta)~\vert~\underset{\mathrm{i} \in \ve{\lambda}_\mathrm{j}}{\cap}\ve{D}_\mathrm{i}\ne\emptyset,~ \vert\ve{\lambda}_\mathrm{j}\vert\geqslant \beta\;.
	\label{eq:lamb}
\end{eqnarray} 

\noindent For example, $\beta=3$ for 2D localization\footnote{note that when the source is mobile, $\beta=2$ will also support 2D localization as shown in \cite{alexandri2018tracking}}, and $\beta=4$ for 3D localization. \remembertext{r1c20}{\new{As shown in Fig.~\ref{fig:zones}(A), for $\beta=3$, there are two sets of 3 receivers $\{r_\mathrm{1},~r_\mathrm{2},~r_\mathrm{3}\}\in \ve{\lambda_\mathrm{1}}$, $\{r_\mathrm{1},~r_\mathrm{3},~r_\mathrm{4}\}\in \ve{\lambda_\mathrm{2}}$. The group containing all the sets is $\{\ve{\lambda_\mathrm{1}},~\ve{\lambda_\mathrm{2}}\}\in\ve{\Lambda}(\beta)$.}} 

For a given $\lambda_j$, we define a \textit{localization area}, $L_\mathrm{j}(\lambda_j,\beta)$, as an area resulting from the intersection of at least $\beta$ detection areas. Formally, 
\begin{eqnarray}
	\ve{L}(\ve{\lambda}_\mathrm{j}, \beta)=\underset{\mathrm{i} \in \ve{\lambda}_j}{\cap}\ve{D}_\mathrm{i}~\vert~ \ve{\lambda}_\mathrm{j}\in\ve{\Lambda}(\beta)\;.
	\label{eq:loc}
\end{eqnarray}

\noindent Denote a \textit{usable area}, $\ve{\mathcal{U}}_\mathrm{j}(\lambda_j,\beta)$, as the set of all positions of the node to be localized, $\bar{p}$, in the localization area, $\ve{L}_\mathrm{j}(\lambda_j,\beta)$, for which "good" localization is attainable. Specifically, the usable area is an area for which $\mathrm{GDOP}(\bar{p}, \ve{L}_\mathrm{j}(\lambda_j,\beta)) \leqslant \alpha$, where $\alpha$ is a scalar value determined by e.g., Table \ref{tab:GDOP}. Formally,
\begin{eqnarray}
	\bar{p} \in \ve{\mathcal{U}}_\mathrm{j}(\lambda_j,\beta) ~|~  \mathrm{GDOP}(\bar{p}, \ve{L}_\mathrm{j}(\lambda_j,\beta)) \leqslant \alpha, ~\bar{p}\in \ve{L}_\mathrm{j}(\lambda_j,\beta)\;.
	\label{eq:use}
\end{eqnarray}

\noindent Our fourth set is the \textit{coverage area}, $\mathcal{C}(\beta)$, denoted as the union of all usable areas, 
\begin{eqnarray}
 \mathcal{C}(\beta)=\underset{\mathrm{j}}{\cup}\ve{\mathcal{U}}_\mathrm{j}(\lambda_j,\beta). 
\end{eqnarray}

We formulate the task of receiver deployment as an optimization problem whose solution is locations $\bar{r}_\mathrm{i}, \ \mathrm{i}=1,\ldots,N$, which maximize the size of the coverage area $\|\mathcal{C}\|$:

\begin{eqnarray}
	\begin{aligned}
		&\argmax_{r_\mathrm{1},r_\mathrm{2},...r_\mathrm{N}}{\mathcal{\| C(\beta)\|}}  \\
		\textrm{s.t.} &~~\|\mathcal{C(\beta)}\cap\mathcal{I}\|\geqslant\rho\|\mathcal{I}\|\;.
	\end{aligned}
	\label{eq:target}
\end{eqnarray} 

Since most applications require the coverage of a given area of interest, $\mathcal{I}$, we constrain the solution such that the resulting convergence area covers at least $\rho$ percent of $\mathcal{I}$. This constraint also allows for the discrimination between a "must-be-covered" area and a "nice-to-have" coverage area. An example of such a consideration is presented in \cite{pickholtz2018habitat} for the monitoring of acoustically-tagged herbivorous fish, \textit{Siganus rivulatus}, close to the shore area. In this case, the movements of the herbivorous fish close to the shallow fringing coral reefs are of interest (the "must-be-covered" area), along with information from other areas along the shore (the "nice-to-have" area).

\subsection{The PDAD Approach}
\subsubsection{Optimal Formalization}
We formulate the problem statement in \eqref{eq:target} as a mixed constraint optimization problem. In accordance with \eqref{eq:target}, let $\ve{\mathcal{I}}, ~\ve{D}, ~\ve{L}, ~\ve{\mathcal{U}}$ and $\ve{\mathcal{C}}$ be 3D matrices representing the area-of-interest, the detection area, localization area, usable area, and the coverage area, respectively. The matrices' rows, columns, and depths represent the quantized x,~ y,~ z grid in Cartesian coordinates, with lower-case letters representing the matrix entries with a resolution set by the bathymetry information (e.g., every 5m). For example, entry $i_{k,~l,~m}$ is the $\mathrm{k^{th}},~\mathrm{l^{th}},~\mathrm{m^{th}}$ entry of $\ve{\mathcal{I}}^{[K\times L\times M]}$. The capability to detect a source located in position $\bar{p}(x=k,y=l,z=m)$ by the $\mathrm{i^{th}}$ receiver located in position $\bar{r}_\mathrm{i}$ is defined by the binary operator

\begin{eqnarray}
	d_\mathrm{k,l,m}(\bar{r}_\mathrm{i}) = \begin{cases}
		1 & ~~\textrm{if}~~P_\mathrm{DT}(\bar{p}(k,l,m),\bar{r}_\mathrm{i}) \leqslant \frac{P_\mathrm{SL}\times P_\mathrm{TL}}{P_\mathrm{NL}}
		\\
		0 & ~~\textrm{otherwise}
	\end{cases}
\end{eqnarray}

Then, a source located in position $\bar{p}$ can be localized if it is detected by at least $\beta$ receivers. The elements of $L(\ve{\lambda}_\mathrm{j},\beta)$ are formalized by the binary operator
\begin{eqnarray}
	l_\mathrm{k,l,m}(\ve{\lambda}_\mathrm{j})= \begin{cases}
		1 & ~~\sum\limits_\mathrm{i=1}^N d_\mathrm{k,l,m}(\bar{r}_\mathrm{i})\geqslant \beta, ~\bar{r}_\mathrm{i} \in \ve{\lambda}_\mathrm{j}\\
		0 & ~~\textrm{otherwise}\;,
	\end{cases}
\end{eqnarray}

\noindent where each $\ve{\lambda_\mathrm{j}}$ is a subset of $\ve{\Lambda}(\beta)$, containing at least $\beta$ receivers. Here, $\ve{\Lambda}(\beta)$ is the complete set of $\bar{r}_\mathrm{i}$ receivers i = 1,\ldots,N. Note that the maximum number of such $\lambda_j$ sets is $J = \sum\limits_\mathrm{p=\beta}^N {N \choose p}$. 

Recall that the usable area $\ve{\mathcal{U}}_\mathrm{j}$ is the mapping of all locations $\bar{p}$ inside $\ve{L}(\ve{\lambda}_\mathrm{j})(\beta)$ whose GDOP is smaller than a given threshold $\alpha$. In the matrix representation,
\begin{eqnarray}
	u_\mathrm{k,l,m}(\ve{\lambda}_\mathrm{j},(\beta))= \begin{cases}
		1 & \ve{\mathrm{GDOP}}(\ve{L}_\mathrm{j}(\lambda_j,\beta)) \leqslant\alpha \\
		0 & \textrm{otherwise}\;.
	\end{cases}
\end{eqnarray}

\noindent Similarly, the coverage area matrix, defined as the union of all the different usable areas $j=\mathrm{1...J}$, is formalized by the binary operator

\begin{eqnarray}
	c_\mathrm{k,l,m}(\beta) = \begin{cases}
		1 & \textrm{if}~~ u_\mathrm{k,l,m}(\ve{\Lambda},(\beta))=1\\
		0 & \textrm{otherwise}\;.
	\end{cases}
\end{eqnarray}

\noindent The intersection of $\mathcal{C(\beta)}$ and $\mathcal{I}$ in the constraint of \eqref{eq:target} can thus be expressed by

\begin{eqnarray}
	\|\mathcal{C}(\beta)\cap\mathcal{I}\| = \sum\limits_{k}\sum\limits_{l}\sum\limits_{m}c_\mathrm{k,l,m}(\beta)i_\mathrm{k,l,m}\;. 
\end{eqnarray}

Note that operators $d_{k,l,m}$, $l_{k,l,m}$, $u_{k,l,m}$ and $c_{k,l,m}$ are all a function of locations $\bar{r}_1,\ldots,\bar{r}_N$, which in turn can take any value within $\mathcal{C}$ and whose determination is the goal of this work. The deployment is obtained by solving

\begin{subequations}
	\begin{align}
		\ve{\Lambda}=\argmax_{\bar{r}_\mathrm{1},\ldots,\bar{r}_\mathrm{N}}&{\sum\limits_{k}\sum\limits_{l}\sum\limits_{m}c_{k,l,m}} \label{eq:subeq1}\\
		\textrm{s.t.}~~&
		\sum\limits_{k}\sum\limits_{l}\sum\limits_{m}c_{k,l,m} 	i_{k,l,m} \geqslant \rho \sum\limits_{k}\sum\limits_{l}\sum\limits_{m}i_{k,l,m} 
		\label{eq:subeq6}	
	\end{align}
	\label{eq:opt}
\end{subequations}

\subsubsection{Algorithmic Solution}
A closed-form solution of \eqref{eq:opt} for the simple case of three receivers deployed over a flat seabed with isotropic propagation loss is presented in the Appendix. We note that, in the general case, problem \eqref{eq:opt} is non-convex. This is because, as illustrated in Fig. \ref{fig:zones}, even when the detection area of each receiver is convex (a circle or an ellipsoid) - and thus so is the localization area - the coverage area may be constructed from a number of non-continuous usable areas. Furthermore, as demonstrated in Figs. \ref{fig:zones}(A) and \ref{fig:zones}(B), even for a convex localization area, the usable area itself may be non-convex. Hence, \eqref{eq:opt} is a constrained non-convex optimization problem, which can be solved by procedures such as branch and bound (B\&B) \cite{clausen1999branch} with a polynomial complexity on average \cite{shabi2017planning,zhang1992average}, or by randomized approaches such as simulated annealing \cite{blum2021learning}. Here, we propose to use GA, which is suitable for complex deployments, i.e., a larger number of receivers and diversified bathymetric and bathythermal conditions. We chose GA since it is suitable for overcoming local minima in problems involving a non-convex objective function \cite{mnasri2015genetic}. The fitness function is multi-objective, seeking to maximize the coverage area inside $\mathcal{I}$ and to minimize a penalty function for coverage outside $\mathcal{I}$.

Evolutionary algorithms (EA) have probabilistic convergence time \cite{ankenbrandt1991extension}. The average convergence time is defined as the number of generations it takes to reach convergence \cite{rylander2001computational}. To that end, the complexity depends on the individual's and population's representation, the implementation of mutation, crossover and selection processes, and the fitness function \cite{corus2017level}. Given the above, the complexity is on the order of $\mathcal{O}(gpi)$, where $g$ is the number of generations, $p$ is the population size, and $i$ is the size of the individuals. \remembertext{r1c23}{\new{We implemented PDAD using Python’s “DEAP” evolutionary computation framework package and used it both for the simulation and the sea experiment.}}

\subsection{Numerical Investigation Setup}

\begin{figure}[h]
	\centering
	\includegraphics[width=\columnwidth]{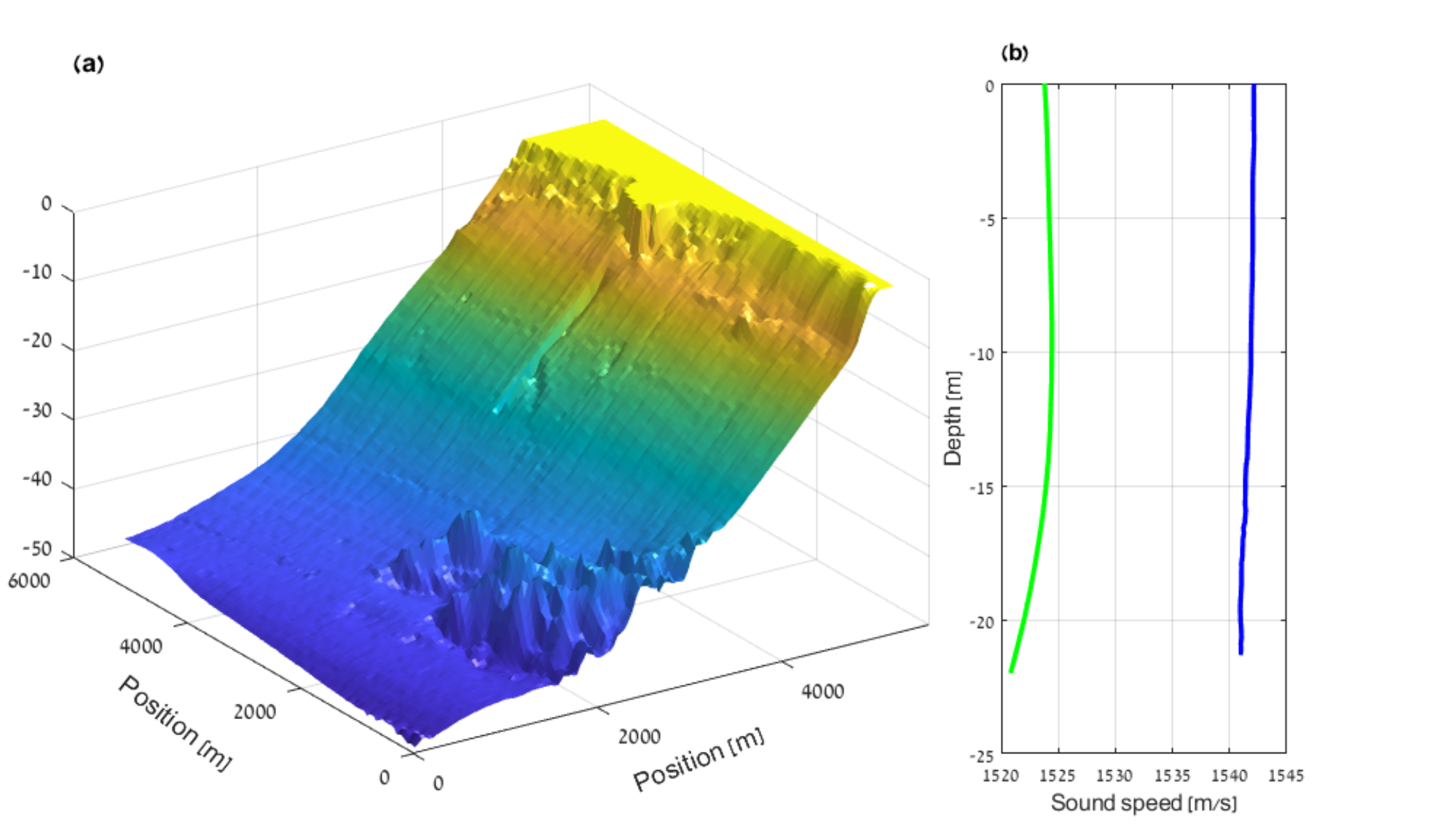}
	\captionsetup{width=.8\linewidth}
	\caption{(a) Bathymetry, (b) eastern Mediterranean winter SV, $\mathrm{SVP_{H}}$ (blue) measured sea experiment $\mathrm{SVP_{M}}$ (green) of the area near ``Orot Rabin'' Power and Desalination Station in Hadera, Israel.}
	\label{fig:SdotYam} 
\end{figure}

\begin{figure}[h]
	\centering
	\includegraphics[width=0.8\columnwidth]{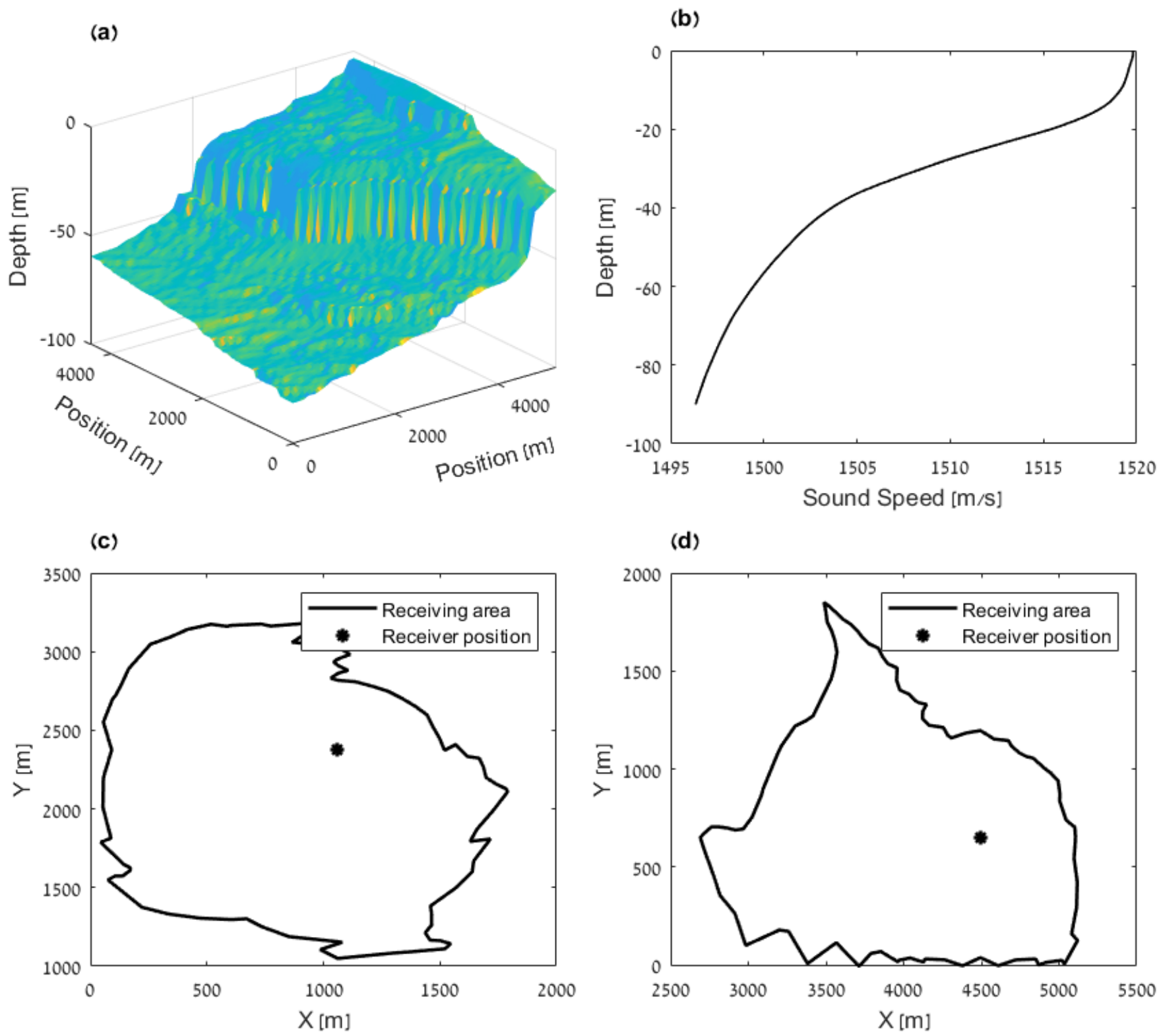}
	\caption{Environmental condition of shallow water close to San Diego Bay. (a) Bathymetry and (b) the sound velocity profile. The impact on the expected detection area of a single receiver is shown in (c) for a receiver positioned at (1000, 2400, -10) and in (d) for a receiver positioned at (4500, 750, -10)}
	\label{fig:SanDiego} 
\end{figure}

To analyze PDAD performance, we consider three environments with different attributes:
\begin{enumerate}
	\item A theoretical, simple environment: a flat seabed with an isotropic SVP, termed $\mathrm{SVP_{ISO}}$. This type of environment may be considered when the environmental conditions are unknown.
	\item A moderate spatially diverse area: an area of $6000 \times 6000 ~\mathrm{m^2}$ shallow water area close to the ``Orot Rabin'' Power and Desalination Station in Hadera, Israel $(32^028'N; 34^052'E)$; an area we also explored in our sea experiment. Within this area, the selected $\mathcal{I}$ area is a $2000 \times 2000 ~\mathrm{m^2}$ rectangle with a water depth ranging from 0 to 25 meters. The entire area was divided into a grid of 60 $\times$ 60, yielding a sample resolution of 100 m. Here, the receivers are anchored at a depth of 0.5~m above the seabed, and the mobile transmitter moves at a depth of 3~m. The bathymetry of this area is shown in Fig.~\ref{fig:SdotYam}(A).
	\item An extremely diverse seabed: a coastal area north of San Diego between $32.65^o$ N to $32.755^o$ N and $-117.265^o$ W and $-117.35^o$ W \cite{divinis2021ngdc}. Out of this $5000 \times 4000$~$\mathrm{m^2}$ area, we picked $\mathcal{I}$ to be a square of $3000 \times 3000 ~\mathrm{m^2}$, which was divided into grid cells of a 100~m resolution. For this environment, the simulated receivers are anchored at a depth of 2~m above the seabed, while the mobile node maintains its depth at 10~m. Fig. \ref{fig:SanDiego}(A) shows the bathymetry of the considered area, and its SVP is shown in Fig \ref{fig:SanDiego}(B). The diversity of the considered area is demonstrated in Fig.~\ref{fig:SanDiego}(C) and \ref{fig:SanDiego}(D), showing significant differences between the effective detection area for two different grid positions.
\end{enumerate}

We consider the specification of an actual acoustic tag manufactured by Thelma-Biotel Inc., Trondheim, Norway \cite{alexandri2018tracking}. These acoustic tags are used globally in applications to monitor fauna in marine environments and to track migration patterns \cite{lennox2021laboratory, reubens2021compatibility}. The mobile node is an acoustic tag (model: ID-HP16) emitting 69kHz single-tone signals of intensity 158dB re 1\textmu Pa~$\mathrm{@}$~1m. A range test we performed showed that the detection distance is 1000m in shallow water with a sandy bottom \cite{alexandri2019localization}.

To measure the GDOP in each deployment setup, the spatial-dependent propagation loss must be accounted for. We consider two ways to attain the propagation loss. The first, assigned only in the case of a flat seabed with an isotropic SVP, applies a transmission loss model of

\begin{eqnarray}
	\mathrm{TL} = 10\log(R)+\alpha R/1000,
\end{eqnarray}

\noindent with R being the transmission range and $\alpha$ the absorption parameter. The result is a transmission power loss of 48dB, where $\alpha = 18~\mathrm{dB/km}$ for 69~kHz \cite{urick1983propagation}. For complex environments, we consider the Bellhop ray-tracing propagation model \cite{porter2011bellhop}. In both cases, the detection area is calculated by setting a limit on the signal-to-noise ratio to be above 10 dB. 

For each of the above three areas, two types of deployment setups were compared. The first is based on the CP method of positioning the receivers at the vertices of equilateral triangles partially covering the inner area of $\mathcal{I}$. The distance between the receivers is set to half of the tags' specified detection range, namely 500~m. The second deployment setup is based on positioning the receivers according to the PDAD. Each individual is a set of the receivers' coordinates $x$ and $y$. For the GA sub-optimal solution, the initial population size is based on a single individual positioned at the center of $\mathcal{I}$, with the rest of the individuals positioned around the center position. The effectiveness of the two methods is compared in terms of the size of the resulting coverage area for each deployment strategy. To this end, we chose the coverage area to be such that for each point inside $\mathcal{I}$, GDOP $\leq5$.

In our simulations, we used an AMD Ryzen\textsuperscript{TM} Threadripper\textsuperscript{TM} 3990X CPU with 128 threads and measured processing time of about 100 msec per thread per each possible deployment setup. To evaluate the complexity of the proposed deployment scheme, we note that the size of a search space to position $N$ receivers in a given area divided into a rectangular grid of $x\times y$  is $(xy)^N$. For example, for a grid size of $\mathrm{100 \times 100}$ and $N=3$, the search space size is $\mathrm{10^{12}}$. Hence, with the full utilization of our server, a complete search of the search space will last more than 24 years. This is because of the complexity of a solution in which brute-force searches the considered area is $\mathcal{O}((xy)^N)$. In our case, using the PDAD approach, the best solution for the flat bottom isotropic propagation environment was achieved in the range of 500 to 2700 GA's generations for the different available number of receivers. \remembertext{r4c22}{\newnew{Using a 64 core, 128GB memory computer, the processing time was 2:30 hours for 3 receivers deployment and 28 hours for 10 receivers deployment.}}

\subsection{Sea Experiment Setup}

The sea experiment aimed to demonstrate the applicability of our simulation results, and to explore the benefit of the proposed PDAD strategy in the realistic case of a low-resolution possibly-mismatched bathymetry map. The experiment took place in November 2021 close to the ``Orot Rabin'' Power and Desalination Station in Hadera, Israel $(32^028'N; 34^052'E)$, in an area of interest of $1200 \times 1400~ \mathrm{m^2}$, at the southeastern part of the same area used in one of our numerical analysis cases (see the bathymetry map in Fig.~\ref{fig:SdotYam}). We used acoustic equipment from Thelma Biotel AS., Trondheim, Norway - specifically, 7 acoustic tags and the standard receivers that decode these emissions. The experiment involved four kayaks, \remembertext{r4c23}{\newnew{each was towing a small buoy 2~m behind its stern. The tags were attached roughly 2~m below the buoy, and maintained their depth using a balancing weight attached below the tags.}}For groundtruthing, each kayak carried a GPS receiver that logged its location throughout the trial. Two clusters of four receivers each were anchored at a depth of 1 meter above the seabed in the explored area. One cluster was positioned according to the CP approach, and the other according to the results of a PDAD calculation using a given bathymetry map of the area and an SVP measured prior to the experiment. The measured SVP, $\mathrm{SVP_{M}}$, is shown in Fig.~\ref{fig:SdotYam}(C).

In addition to comparing the size of the usable area for CP and PDAD, a second performance metric explored the \remembertext{r4c25a}{\newnew{throughput of the tags - receivers link, defined as $\xi_{i, r}$. Here, the throughput is defined by the ratio between the number of receptions to the total number of emissions. Specifically, two cases are considered. 1) $\xi_{2, r}$ is the ratio between the number of received emissions by one or two receivers and the total number of emissions, and 2) $\xi_{3, r}$ it the ratio between the number of received emissions by three or four receivers and the total number of emissions throughout the experiment. The former reflects on the detection properties of the deployed setup, while the later on the localization quality.}} The throughput metric accounts for possible uneven time spent by the kayaks in the CP or PDAD setups. To avoid bias, we normalized the throughput by the tags' distances to the center of each area. Specifically, for each cluster of four receivers, the geometric center of the area was calculated. Then, the throughput was calculated and normalized by the tag's range from the cluster centroid.

Fig. \ref{fig:exp_cover} shows the receivers' locations in the explored area. The position of the CP's receivers are marked by aqua-colored squares, and the positions of the PDAD's receivers are marked by red-colored diamonds. In a previous work \cite{alexandri2019localization}, a range test showed that in the considered environment, the detection range of a similar tag and receiver pair is roughly 1,000 m. Thus, for CP, the distance between the receivers was set to 500m. We note that both CP and PDAD shared receiver number 2.

The experiment lasted for 240 min. During that time, each tag emitted a signal in a fixed interval every 30 to 45~sec, for a total of 2,730 emissions. Out of these, 406 emissions were detected by at least one receiver and 280 emissions were detected by at least three receivers of the CP or PDAD clusters. \remembertext{r4c27a}{\newnew{In order to ensure that the four kayaks cover the complete area of interest, their route, provided in Fig. \ref{fig:exp_cover}, was planned to reach beyond the anticipated detection range of the tags. As a result, we report low tag detection rate.}} For each detected emission, the receivers measured the ToA by their internal clock. The receivers were time-synchronized prior and after the experiment. This involved both synching the receivers' clock to a reference clock by attaching a specific acoustic tag to one of the receivers, and using its emissions to time-synchronize the others (see details in \cite{alexandri2018tracking}). Tag emissions were sorted and aggregated to the CP and PDAD clusters. If received by either three or four receivers, the emissions were considered to be inside the usable area that allows localization. \remembertext{r4c42a}{\newnew{To comment on the receiving conditions we also recorded the ambient noise level, as measured by each receiver.}}

\section{Results}
In this section, we report results from the numerical investigation of our deployment method and from the designated sea experiment. For the simulations, we used the size of the area which achieves good localization, i.e., $\mathrm{GDOP} \leqslant 5$, and explored the results against the brute-force solution and compared to CP. We also investigated the sensitivity to the system's parameters - in particular, the number of receivers, the size of $\mathcal{C}$ for different GDOP values, and to different SVPs. For reproducibility, we shared our implementation code in the supplementary material.

\subsubsection{Simulation results} \label{sec:sim_res}    
\begin{figure}[t]
	\centering
	\includegraphics[width=\columnwidth]{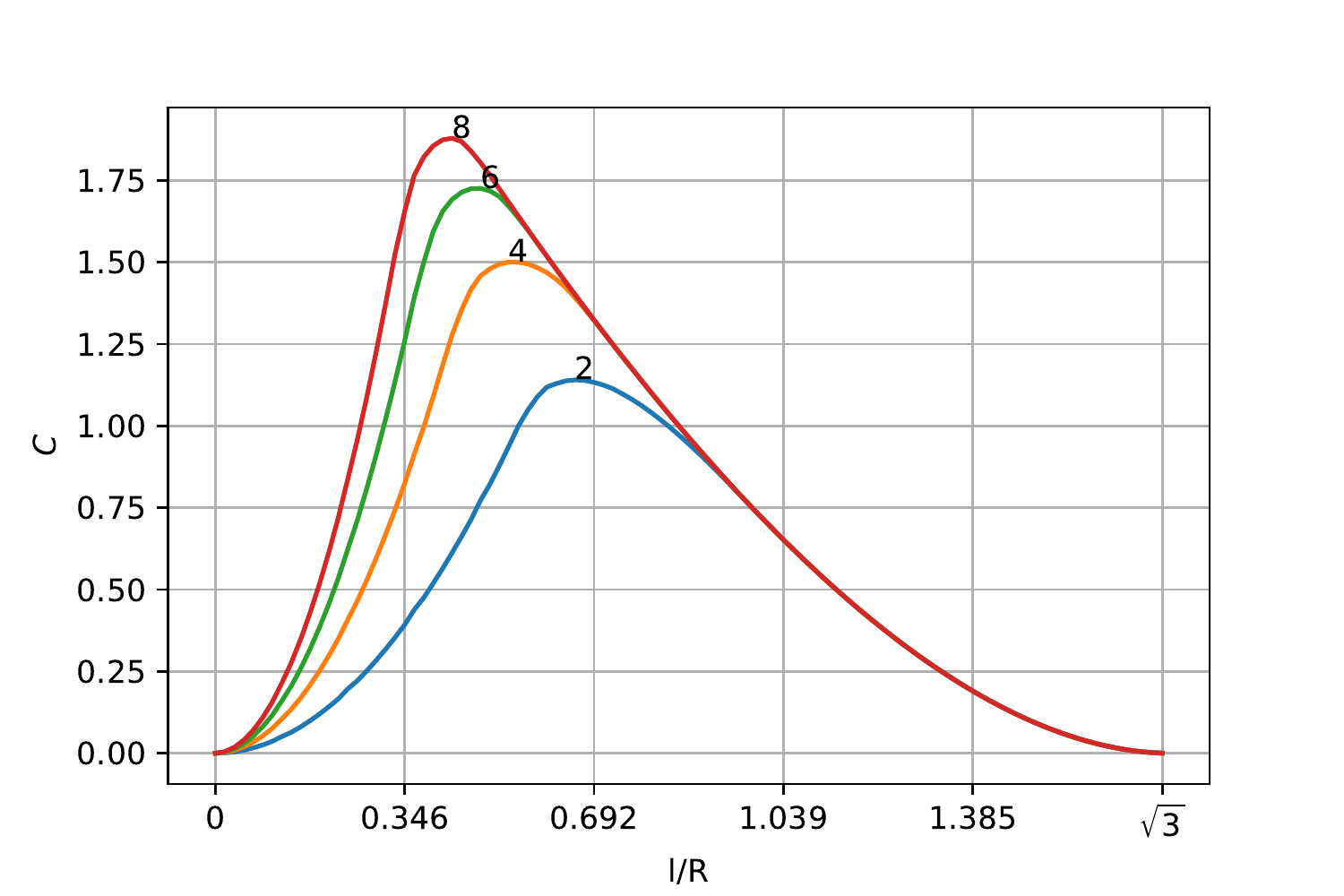}
	\captionsetup{width=.8\linewidth}
	\caption{Effect of distance between the receivers, $l$, normalized to the maximum receiving range, $R$, on covered receiving area, $\mathcal{C}$, and attained positioning quality for 3 equilateral receivers' deployment in isotropic propagation loss. As the quality of positioning increases (the GDOP is smaller), the attained covered area $\mathcal{C}$ decreases.} 
	\label{fig:toy}
\end{figure}

We start by analyzing the results of the theoretical environment. Fig.~\ref{fig:toy} shows the size of  $\mathcal{C}(\beta=3)$ for the case of three receivers as a function of the distance between the receivers, $l$, normalized by their receiving range, $R$. The maximum coverage area for GDOP=5 is attained when the distance between the receivers is about half of the receiving range. For cases of higher GDOP values, i.e. the localization accuracy in some parts of $\mathcal{C}$ is of lower quality, a larger coverage area can be attained. For example, for GDOP=8, the optimal distance between the receivers is about 40\% of the receiving range and $\mathcal{C}=1.85R^2$. \remembertext{r4c31a}{\newnew{For a lower GDOP, when localization accuracy is of high priority, a smaller coverage area is attainable. For example for $\mathrm{GDOP}=2$ the maximal attainable coverage area is $\mathcal{C}=1.14R^2$ for $l= 0.66$. This result may serve as a guideline for deployments in non-spatially diverse areas.}} The analytical derivation of the attainable coverage area for the theoretical environment is outlined in the Appendix. In Fig. \ref{fig:com_vs_ga}, we compare the achieved coverage area between CP and PDAD for different numbers of receivers, ranging from 3 to 13. We observe that the coverage area gained by using PDAD over CP increases with the number of receivers. This is attributed to the increase in the number of degrees of freedom for the receivers' placement.       

\begin{figure}[t]
	\centering
	\includegraphics[width=\columnwidth]{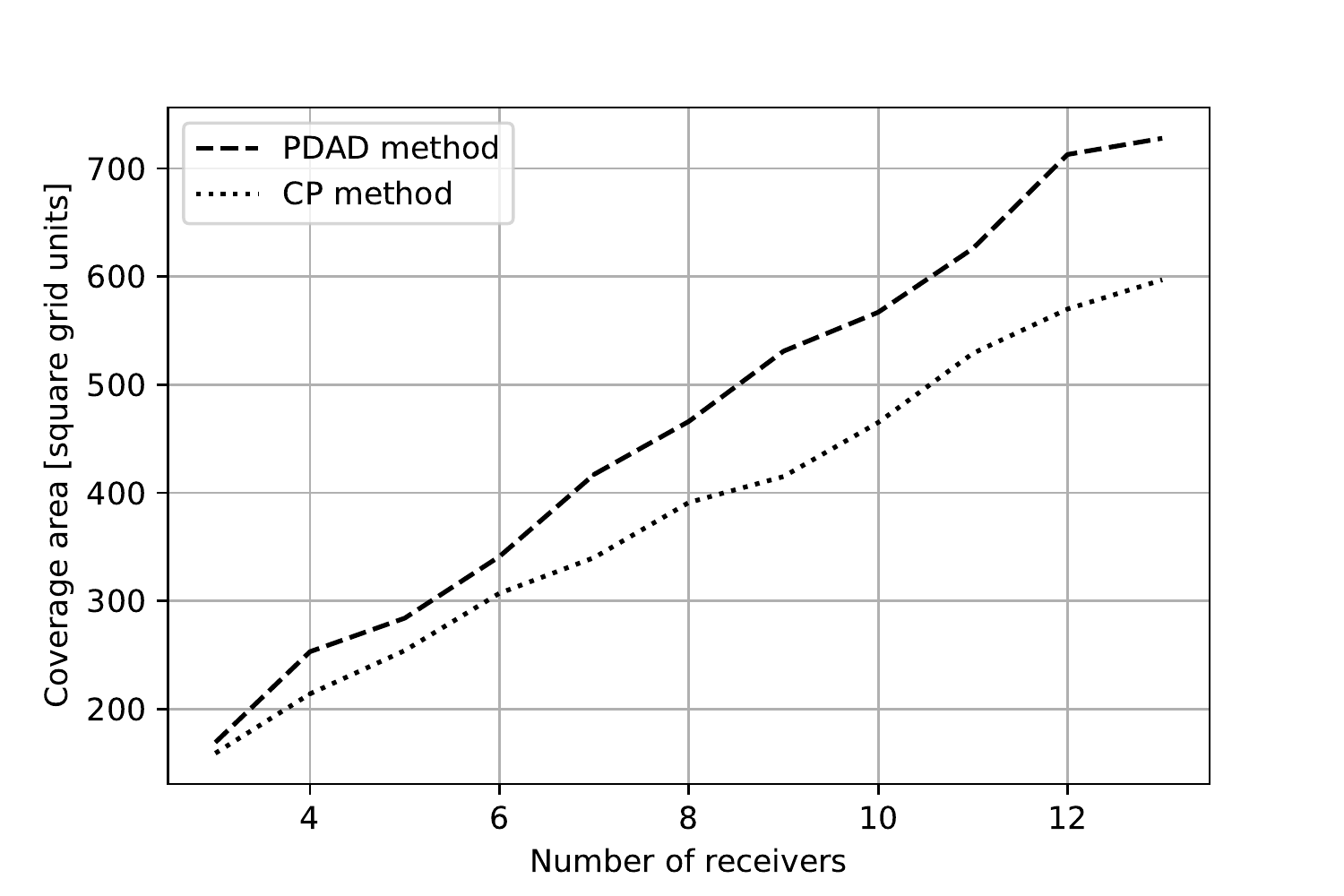}
	\captionsetup{width=.8\linewidth}
	\caption{Area covered by deployment setup based on CP method, i.e., positioning the receivers at the vertices of equilateral triangles vs. deployment based on the proposed PDAD method, i.e., optimization of receivers locations based on the surrounding environmental conditions.} 
	\label{fig:com_vs_ga}
\end{figure} 
For the moderate spatially diverse area, in Table~\ref{tab:dep_1}, we compare the results for two SVPs: a simple fixed profile of 1520~m/s, $\mathrm{SVP_{ISO}}$, and an eastern Mediterranean winter SVP, as shown in Fig.~\ref{fig:SdotYam}(B) \cite{salon2003sound}, termed $\mathrm{SVP_{H}}$. The first two columns of Table \ref{tab:dep_1} summarizes the ratio between the size of the coverage area as obtained by PDAD and CP, for the moderate diverse bathymetry in Hadera

\begin{eqnarray}
	\eta=\mathcal{C}_\mathrm{PDAD}/\mathcal{C}_\mathrm{CP}.
	\label{eq:ratio_1}
\end{eqnarray}  

\begin{figure}[t]
	\centering
	\includegraphics[width=0.6\columnwidth]{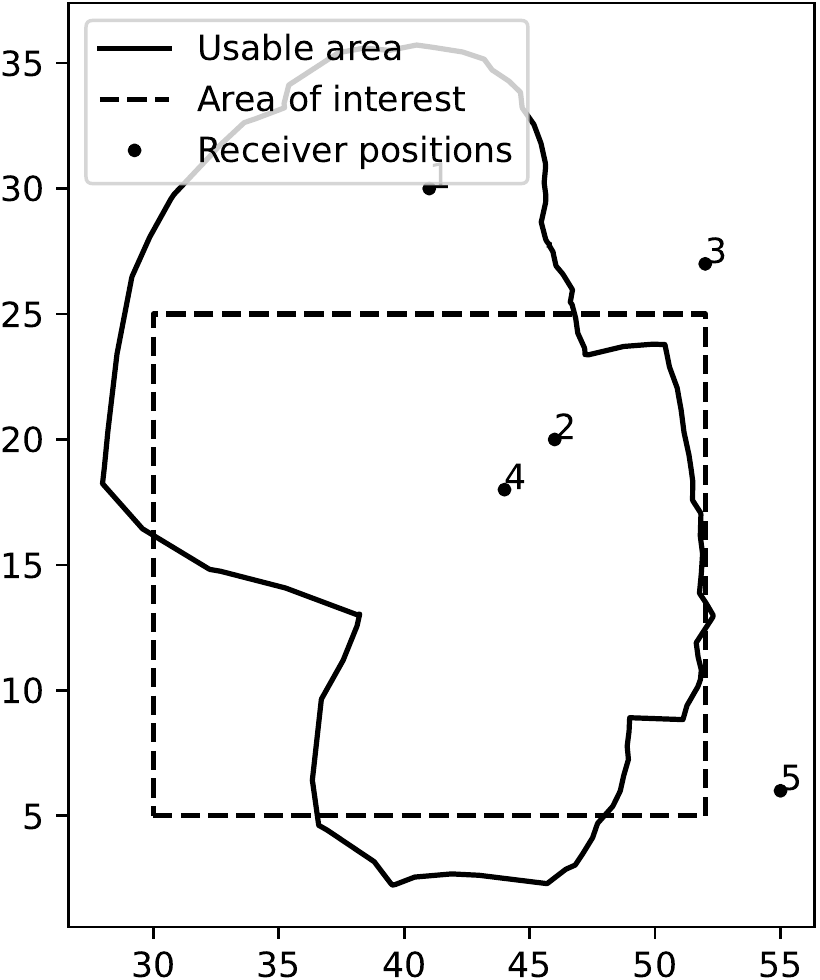}
	\caption{Usable area for the deployment of 5 receivers in isotropic SVP, $\mathrm{SVP_{ISO}}$, in a moderate spatially diverse area. Note that 3 out of the 5 receivers are positioned outside of $\mathcal{I}$.} 
	\label{fig:iso_5}
\end{figure}

\noindent The results are presented in Table~\ref{tab:dep_1} for the different number of receivers, and for GDOP$\leq5$. We observe that, using PDAD, for both SVPs the attainable covered area is larger than that of CP. An interesting result presented in Fig.~\ref{fig:iso_5} shows that for the deployment of 5 receivers in isotropic conditions - in contrast to CP where all of the receivers are placed inside $\mathcal{I}$ - the PDAD solution suggests that 3 out of the 5 receivers be placed outside $\mathcal{I}$. This deployment yields an increase of 67\% in the localized area. Finally, from the results in Table~\ref{tab:dep_1}, we observe the difference of the gain obtained for the two different SVPs. A much higher gain in using PDAD is shown when the SVP is complex. This is because a diverse SVP impacts the propagation loss, rendering the channel to be spatial dependent.

\begin{table}[h]
	\caption{$\eta$ - from \eqref{eq:ratio_1} for deployment of 3, 5 and 10 receivers at the Hadera power station and in San Diego. The results are shown for Hadera's $\mathrm{SVP_{H}}$ and a theoretical isotropic $\mathrm{SVP_{ISO}}$, and for San Diego's $\mathrm{SVP_{S}}$ and theoretical isotropic $\mathrm{SVP_{ISO}}$ for GDOP$\leq$5.}
	\begin{center}
		\begin{tabular}{ccccc}
			\toprule 
			&  Hadera &  Hadera  & San Diego & San Diego\tabularnewline
			&$\mathrm{SVP_{H}}$ & $\mathrm{SVP_{ISO}}$ & $\mathrm{SVP_{S}}$  & $\mathrm{SVP_{ISO}}$\tabularnewline
			\# of receivers&$\eta$&$\eta$ &$\eta$ &$\eta$\tabularnewline
			\midrule
			3 & 7.3   & 2.8 & 1.6 & 2.33\tabularnewline
			\midrule
			5 &  7.1 & 2.8 & 1.9 & 2\tabularnewline
			\midrule
			10 & 3.5 & 2.02 & 2.3 & 1.74\tabularnewline
			\bottomrule
		\end{tabular}
	\end{center}
	\label{tab:dep_1}
\end{table}

The third and fourth columns of Table \ref{tab:dep_1} show the values of $\eta$ from \eqref{eq:ratio_1} for the third explored environment with the highly diverse bathymetry (the San Diego area) for its $\mathrm{SVP_{S}}$ and a theoretical isotropic $\mathrm{SVP_{ISO}}$. We observe that the size of the coverage area obtained by PDAD is also significantly larger than that of CP in the case of complex bathymetry.  We note that the marginal added coverage area for the SVP in San Diego increases with the number of receivers but an opposite trend is shown for the Hadera area. This is due to the higher diversity of the San Diego seabed and due to the deployment strategy. The former is attributed to bathymetry complexity. That is, when the bathymetry is highly complex, e.g., the one from San Diego, adding receivers helps cover shadow zones and their proper deployment location becomes more important. In less complex environments, e.g., the one from Hadera, adding receivers helps CP cover more area for localization and the gain in using PDAD decreases. We observe that the performance gain in the Hadera area increases for the more complex SVP, but the performance gain in the San Diego area is higher for the isotropic SVP. We explain this by the randomness of the CP method. In particular, in the center of the area of interest, CP may or may not achieve good performance. Still, since PDAD seeks to maximize the convergence area, performance gain is still above 1 in all cases.   

Next, we explore the ratio between the coverage and the usable areas,

\begin{eqnarray}
\theta = \mathcal{C/U}
\label{eq:ratio_2}
\end{eqnarray}  

\noindent for two values of localization qualities, GDOP=5 and GDOP=12. Results for the moderate and complex environments are shown in Table \ref{tab:dep_3} for 3, 5, and 10 receivers. As expected, the results show that for the complex environment, sacrificing the positioning quality, e.g., GDOP=12, may increase the coverage area compared to that of the moderate environment. We argue that this is because of the diverse bathymetry, which impacts the propagation loss, and thereby the channel's spatial diversity.

\begin{table}[h]
	\caption{Ratio between the coverage and the usable area, $\theta$, for the deployment of 3, 5 and 10 receivers for San Diego and the Hadera power station areas, for GDOP = 5 and GDOP = 12. }
	\begin{center}
		\begin{tabular}{ccccc}
			\toprule 
			Environment  & \multicolumn{2}{c}{San Diego} & \multicolumn{2}{c}{Hadera }\tabularnewline
			\midrule
			\# of receivers&\multicolumn{2}{c}{$\theta$}&\multicolumn{2}{c}{$\theta$}\tabularnewline
			for GDOP  &  5 &  12 & 5 &12   \tabularnewline
			
			\midrule
			3 &0.75 &0.94& 1.0& 1.0  \tabularnewline
			\midrule
			5 & 1.0 & 1.0& 0.87 &1.0\tabularnewline
			\midrule
			10 & 0.84 &1.0 &0.99&0.99\tabularnewline
			\bottomrule
		\end{tabular}
	\end{center}
	\label{tab:dep_3}
\end{table}

\begin{figure}[h]
	\centering
	\includegraphics[width=0.9\columnwidth]{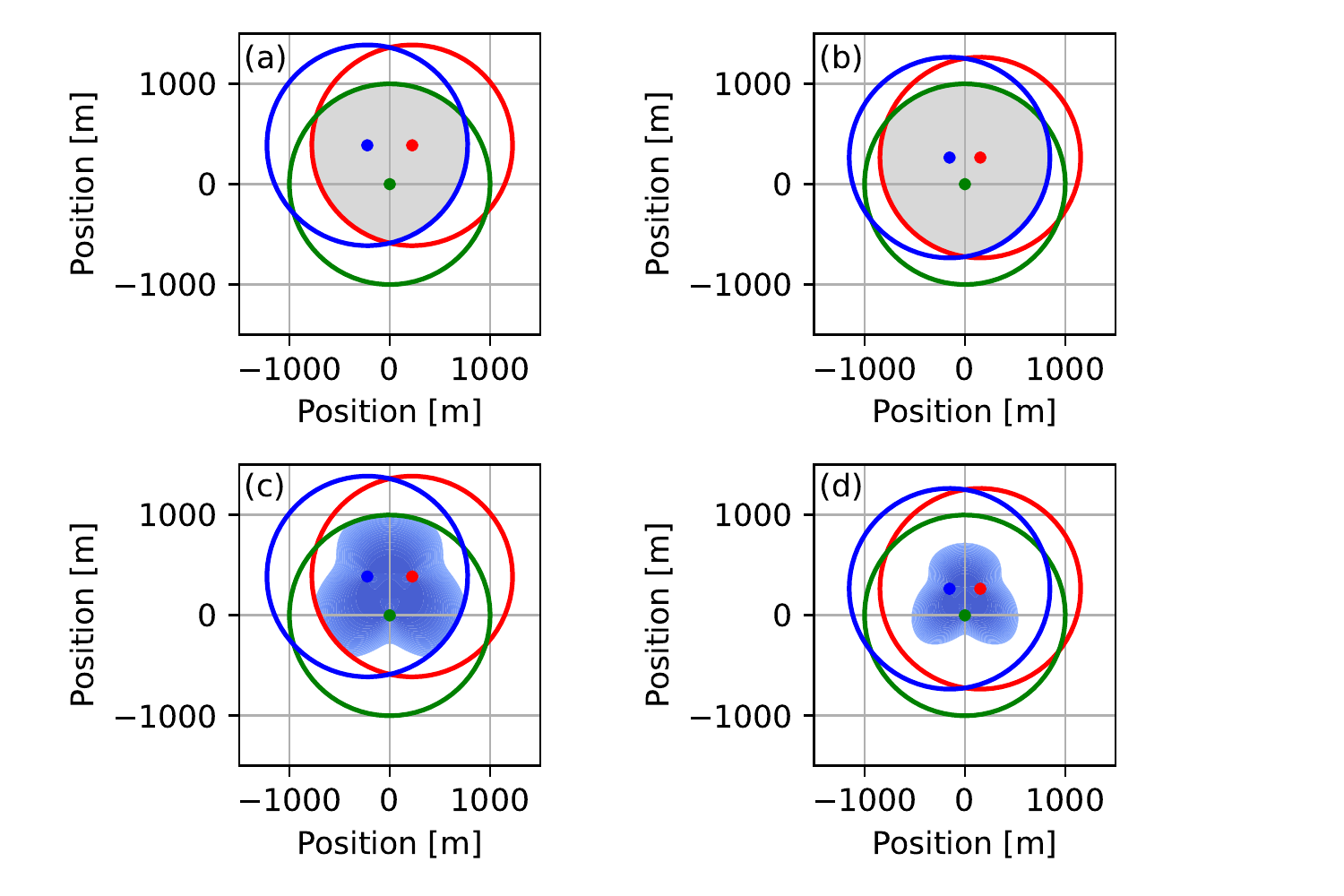}
	\caption{Effect of distance between the receivers on the usable area and the coverage area with attained positioning quality. In setup 1, plots (a) and (c), the receivers are placed $450~\mathrm{m}$ away from each other creating a usable area of $\mathcal{S} \approx 1.782~\mathrm{km}^2$ and the coverage area is $\mathcal{C} \approx 1.544~\mathrm{km}^2$, occupying $86\%$ of the usable area. In setup 2, plots (b) and (d), the receivers are placed $300~\mathrm{m}$ away from each other creating a usable area of $\mathcal{S}\approx2.1472~\mathrm{km}^2$. For this setup, the receivers are closer to each other than in Setup 1, and the coverage area is $\mathcal{C} \approx0.791~\mathrm{km}^2$, occupying only $37\%$ of the usable area. The usable area in plots (a) and (b) is highlighted in gray. The coverage area, plots (c) and (d) for GDOP$\leqslant$5 is highlighted in light blue.}
	\label{fig:demo} 
\end{figure}

The impact of the geometrical relations between 3 receivers on the usable area, $\mathcal{U}$, and on the coverage area, $\mathcal{C}(\beta)$, is demonstrated in Fig. \ref{fig:demo} for $\beta=3$ and $\beta=4$. The green, blue, and red circles are the detection areas of the green, blue, and red receivers positioned at the corresponding colored markers, respectively. Comparing Figs.~\ref{fig:demo}(A) and \ref{fig:demo}(B), we observe that, in the latter, the receivers are located closer to each other and the size of the usable area, $\mathcal{U}$, highlighted in gray, is larger. However, the resulting coverage area, $\mathcal{C}$, highlighted in light blue, for $\alpha=5$, is smaller in Fig. \ref{fig:demo}(D). This is due to the small distance between the receivers, which yields a smaller angle between pairs of receivers and the node to be localized, thereby leading to poor localization.


\subsubsection{Experiment results }
\begin{figure}[ht]
	\centering
	\includegraphics[width=\columnwidth]{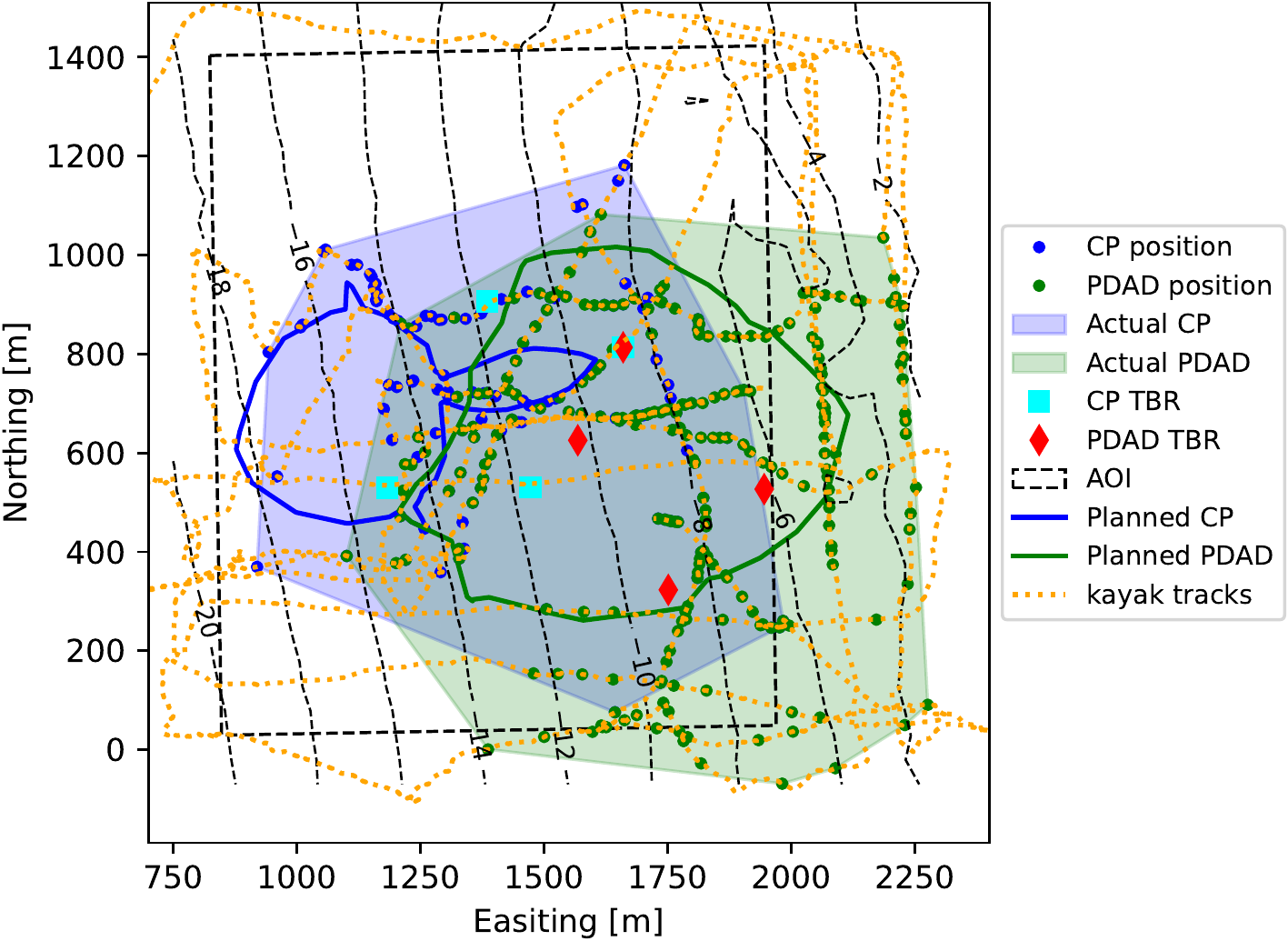}
	\captionsetup{width=.8\linewidth}
	\caption{Positioning of tags and coverage areas for CP and PDAD methods. Light blue and green-filled areas are the minimum convex hull encompassing all points of CP and PDAD, respectively. Solid lines are the planned usable areas. Triangles are the receivers' positions. Blue and green dots are emissions received by at least three receivers of the CP and PDAD sets, respectively.}
	\label{fig:exp_cover} 
\end{figure}

As our first performance metric, we explore the size of the usable area for CP and PDAD. Fig. \ref{fig:exp_cover} shows the position of tags received by at least three receivers for each deployment strategy. For the sake of comparison, we also show the planned usable area and the minimum convex hull for each method. We observe that the size of the usable area obtained by PDAD is 30\%  larger than that produced by CP. We also observe that, in the case of CP, three of the receivers are positioned outside the usable area. This is because CP does not account for the area's specific propagation conditions. Finally, we note that compared to the planned usable area of the CP and PDAD - marked by solid blue and green lines, respectively - the actual usable areas span to the east further than expected. This is attributed to the acoustic propagation conditions on the day of the experiment, which were likely better than those at the time of the tag's range testing. 

\begin{figure}[ht]
	\centering
	\includegraphics[width=\columnwidth]{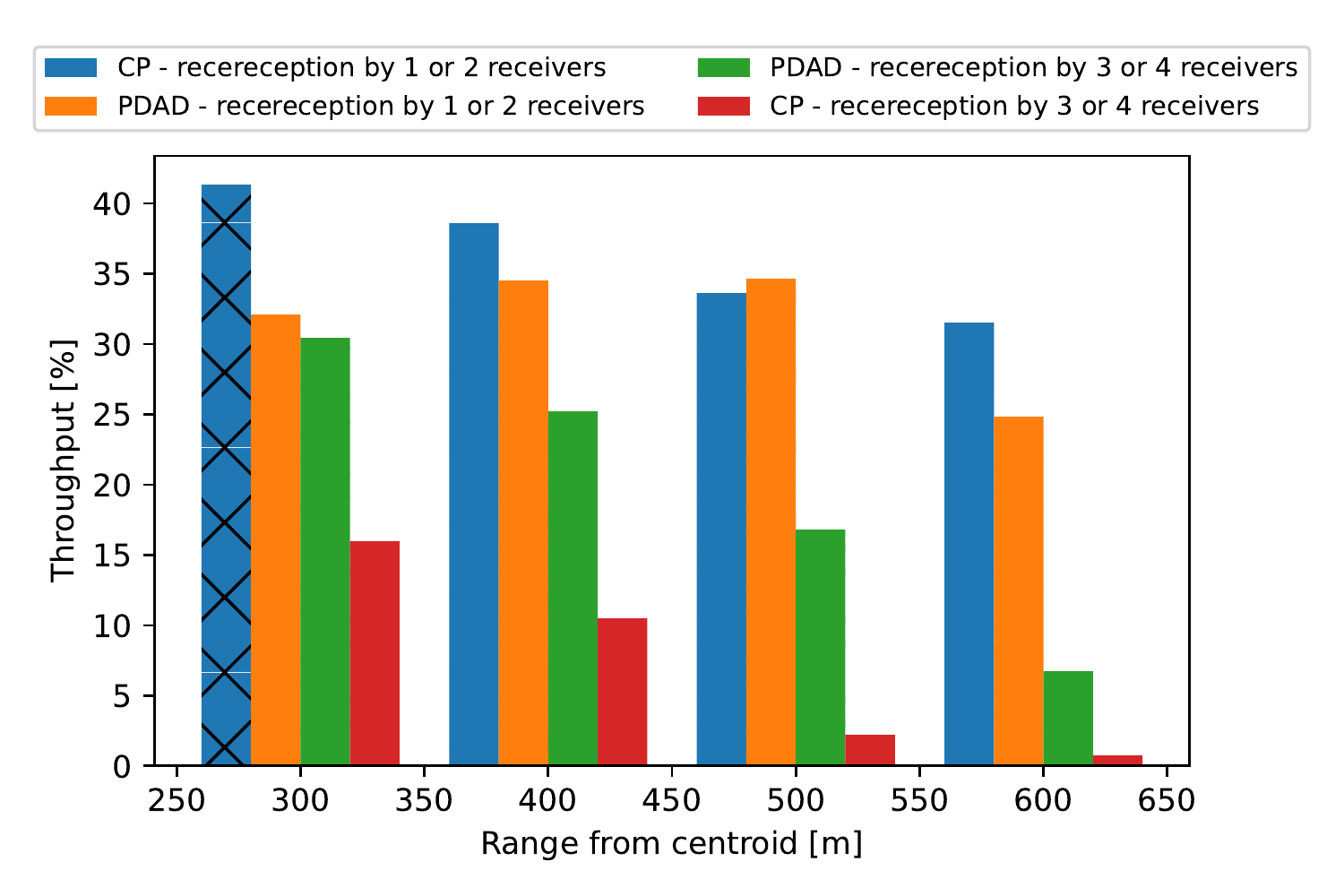}
	\captionsetup{width=.8\linewidth}
	\caption{$\xi_{2, r}$ - ratio between emissions received by one or two receivers and the total number of acoustic emissions (CP - blue, PDAD - orange), and $\xi_{3, r}$ - ratio between localized tags to the total number of acoustic emissions (CP - red, PDAD - green). For the PDAD method, the throughput of localized emissions is better than the CP as the range increases, i.e., more emissions are localized at larger ranges for the PDAD method, leading to a larger usable area.}
	\label{fig:thoughput} 
\end{figure}

As a second performance metric, we explored the throughput of the tags - receivers link. The results presented in Fig. \ref{fig:thoughput} show that the performance benefit of PDAD over CP in terms of the throughput increases with the above range. This implies that tags farther away from the receivers can still be localized by the PDAD cluster; thus, the usable PDAD area is effectively larger than that of CP. The figure also shows the throughput performance for less than three detecting receivers. While this setup does not allow localization without ambiguation, such detection indications can still provide valuable positioning information and is thus of interest \cite{baktoft2017positioning}. We note that, for emissions received by three or four receivers, the PDAD's throughput is higher than that of CP by at least 80\%. However, for emissions received by one or two receivers the CP throughput is sometimes higher than that of PDAD. This is because the PDAD locations are planned for localization by at least three receivers. Still, the difference is not greater than 25\%, implying that the benefit of gaining larger coverage by PDAD is not highly diminished by the reduction of valuable positioning information. 

\remembertext{r4c42b}{\newnew{In Table \ref{tab:noise}, we report the 10th, 50th and 90th percentiles of the noise measurements as obtained by each receiver. We note the relatively small noise level difference between the 10th and 90th percentiles and conclude that the detection conditions during the experiment did not vary much. From Table \ref{tab:noise}, we further note that two receivers of the PDAD cluster, namely receivers \#~557 and \#~558, experienced higher noise level since their position was close to the shoreline. Still, regardless the higher noise level, the PDAD outperformed CP.}} \remembertext{r4c42c}{\newnew{Finally, we observe from Fig. \ref{fig:exp_cover} that, for PDAD, the tag's detections are consistently along the kayak routes, whereas, for CP, the detections appear more sporadically along these tracks. This is an indication of the better stability in detection that the PDAD setup can obtain.}} 

\begin{table}[h]
	\caption{10 Minutes average noise measured by the receivers.  }
	\begin{center}
		\begin{tabular}{ccccc}
			\toprule 
			&& \multicolumn{3}{c}{Percentile} \tabularnewline
			Cluster & Receiver \#&0.1&0.5&0.9\tabularnewline
			\midrule
			CP& 417 &  8.4 &  9 & 9   \tabularnewline
			\midrule
			CP& 418 &11 &14& 15  \tabularnewline
			\midrule
			CP& 1765 &7.8 & 10& 10.6 \tabularnewline
			\midrule
			CP& 1766 &8.4 & 11& 12 \tabularnewline
			\midrule
			CP \& PDAD & 721 &7.8 & 11 & 13 \tabularnewline
			\midrule
			CP \& PDAD & 1153 &7.4 & 9& 9.6 \tabularnewline
			\midrule
			PDAD & 557 &18 & 23& 29.6 \tabularnewline
			\midrule
			PDAD & 558 & 23 &26 &28  \tabularnewline
			\midrule
			PDAD & 1154 &8 & 9& 10 \tabularnewline
			\midrule
			PDAD & 1155 &9 & 11& 13 \tabularnewline
			\bottomrule
		\end{tabular}
	\end{center}
	\label{tab:noise}
\end{table}

\section{Discussion}		

\remembertext{r4c32}{\newnew{We start our discussion by surveying some relevant solutions to the deployment problem.} The optimal receiver deployment problem} is shown to be NP-Hard with similarities to the k-vertex problem \cite{tekdas2010sensor, garcia2019approximation}. For a given graph G(V, E) and for parameters k, l, this problem seeks to discover whether G contains k vertices that cover at least one edge. Considering the high complexity required to solve the deployment problem optimally, some heuristic solutions are proposed. Much like our sub-optimal GA-based solution for PDAD, these include GA to determine the best locations for the receiving nodes \cite{diez20193d,diez2019genetic}, and \cite{diez2020optimized} by iteratively improving the position of the receivers along a given grid, while maintaining constraints in the form of clock drifts and the existence of obstacles. Due to its diversification and intensification in the search within the space of solutions, GA can also avoid falling into local minima \cite{mnasri2015genetic} for diverse propagation loss conditions \cite{diez2019genetic}. Another approach is to compute a Pareto front with a diversified local search for the optimal placement of nodes \cite{roa2007optimal}. Here, the deployment plan is examined under two local criteria: reception availability and quality of positioning. A different approach is simulated annealing, where a set of n nodes to be localized are randomly selected and the positions of the anchor nodes are stochastically optimized to increase the accuracy of the localization estimation \cite{niewiadomska2009optimization}. Other heuristics are particle swarm optimization \cite{wang2007distributed} and Tabu search methodologies \cite{laguna2009diversified}. The firefly algorithm \cite{tuba2018two} is used to initialize anchors' positions at the corners of the explored area, and then move them trying to increase the angle between anchors and the node to be localized, while the relative distance between the anchors and the node to be localized is decreased.

The above works obtain good results for terrestrial networks. Yet, some of the underlying assumptions may be too hard for the underwater acoustic environment. Specifically, it is assumed that the propagation loss does not change with space; that the node to be localized is either bounded inside a polygon whose anchors are its vertices or that some information about its route is known; and that a receiver can be added upon demand to increase the quality of the covered area. In our considered case, due to the spatially-dependent bathymetry, the propagation loss is often a complex location-dependent function, and the optimum coverage solution dictates positioning of receivers in non-overlapping areas.

\subsection{Using the GDOP to Measure Localization}
\newnew{We would like to comment about our usage of the GDOP (cf. \cite{sharp2009gdop}) as a utility metric.} We use the GDOP to assess the expected accuracy of the localization by the receivers' deployment setup. Other common metrics are the Cram\'{e}r-Rao lower bound (CRLB), which expresses a lower bound on the variance of unbiased estimators, and which is widely used as a positioning performance estimator in wireless networks \cite{miao2007positioning}. Alternatively, the root mean square error (RMSE) of the location estimates or the cumulative distribution probability of the location errors can both be used as a localization quality of measures \cite{sharp2009gdop}. However, it is useful to decouple the statistical error component of the positional error from the geometric factors of the deployment setup. \newnew{We find this representation in the GDOP metric.} The GDOP quantifies how errors in the ToA measurements translate into the covariance components of the estimated position, and represents the influence of the standard deviation of the measurement errors onto the solution. The GDOP values can be categorized to measure the localization quality. One such option is presented in Table \ref{tab:GDOP} as proposed in \cite{dutt2009investigation}.

\begin{table}[ht]
	\caption{GDOP Ratings}
	\begin{center}
		\begin{tabular}{cl}
			\toprule 
			GDOP Value $(\alpha)$ & Ratings\tabularnewline
			\midrule
			1 & Ideal\tabularnewline
			2-4 & Excellent\tabularnewline
			4-6 & Good\tabularnewline
			6-8 & Moderate\tabularnewline
			8-20 & Fair\tabularnewline
			$\geqslant$20 & Poor\tabularnewline
			\bottomrule
		\end{tabular}
		\par\end{center}
	\label{tab:GDOP}
\end{table}

\subsection{Discussion of Results}
\newnew{We next make some comments about our results.} \remembertext{r1c29}{\new{As demonstrated in our sea experiment, optimized placement of the receivers can gain extra coverage area of about $30\%$. This is attributed to the high dependency of the underwater acoustic propagation on environmental conditions, such as temperature profile and the bathymetric map. While such a diversity may also be present in terrestrial or aerial testbeds, it is highly dominant in the underwater acoustic environment with spatial variations on the order of a few tens of meters. Hence, while our method can be applied also for other domains, it is mostly attractive for underwater testbeds. Furthermore, as the tag's size determines the detection range, we target our method to the tracking of marine megafauna. This is because for fish or small animals like Lobsters, the detection range of the tags \remembertext{r4c35a}{\newnew{in shallow water/noisy environments}}is on the order of a few tens of meters (see \cite{alexandri2018tracking}), whereas for	megafauna like sharks on which large tags can be mounted, the detection range is on the order of km \cite{alexandri2019localization} \remembertext{r4c35b}{\newnew{even in shallow/noisy water.}}\remembertext{r4c34}{\newnew{Since the affect of spatial diversity of the acoustic propagation becomes more dominant at scales of tens of meters, which is for many cases, beyond the detection range of small tags, the method is mostly useful for the larger tags - i.e., for the task of tracking megafauna.}} Hence, by using our method, the 30\% increase in the explored area holds much more impact for the task of tracking the locations of marine megafauna.}}
 
\subsection{Conclusions}\label{sec:concl}
This study focused on developing a systematic framework for planning the deployment of underwater receivers for the task of localizing acoustically-tagged marine megafauna such as sharks, sea turtles, and seals. We formalized the deployment position as a constraint optimization problem that takes into account the environmental conditions, the desired localization quality, and a given area-of-interest that should be covered with high priority. \remembertext{r4c31b}{\newnew{For a flat bottom bathymetry and isothermal conditions, we showed that the common practice can achieve optimal coverage area.}}For the case of complex bathymetry where the complexity is too high to directly solve the problem, we offered a sub-optimal solution based on a genetic algorithm that is able to efficiently solve the problem for large areas of a few square kms and for a large number of receivers. We explored the benefits of our proposed approach in terms of the size of the coverage area and the throughput of the tags' emissions. The results are presented for the numerical simulations and \remembertext{r1c28}{\new{verified in a field experiment, comparing the expected analytical results to \textit{in-situ} measurements,}} and show that the performance of our proposed deployment - in terms of the coverage area - is superior to that of the common practice. Future work will further investigate how the setup can better account for the expected seasonal changes of the SVP.

\section{Data Availability Statement}
Code for the deployment setup framework with a sample dataset can be found at \href{https://github.com/kerentalmon/Receiver-Deployment.git}{GitHub}\footnote{\url{https://github.com/kerentalmon/Receiver-Deployment.git}}. The dataset collected during the experiment (time-of-arrival indications from all receivers, and GPS locations of the kayaks and receivers) is provided in the \href{https://figshare.com/articles/dataset/Tags_emissions/17942534}{Supplementary Material section}\footnote{\url{https://figshare.com/articles/dataset/Tags_emissions/17942534}} \cite{Alexandri2022}. Further inquiries can be directed to the corresponding author.

\section{Author Contributions}
TA conceived and led the study. TA and RD conducted the field work. TA analyzed the data. Both authors interpreted the findings, wrote the manuscript, and RD approved the final version.

\section{Funding}
This work was supported in part by the MOST-BMBF German-Israeli Cooperation in Marine Sciences 2018-2020 (Grant \# 3-16573), and by the MOST action for Agriculture, Environment, and Water for the year 2019 (Grant \# 3-16728).

\newpage

\section*{Appendix} 

An interesting case to explore is the deployment of 3 receivers on a flat seabed with isotropic propagation loss and with no constraint on the area of interest. The question explored is: What is the best receivers setup to attain the maximal covered area $\mathcal{C}$~? The  expected answer is an equilateral triangle setup \cite{fewell2006area,shiu2010divide}. Yet, the proof is not trivial .

\begin{figure}[h]
	\vspace{-0.5cm}
	\centering
	\includegraphics[width=\columnwidth]{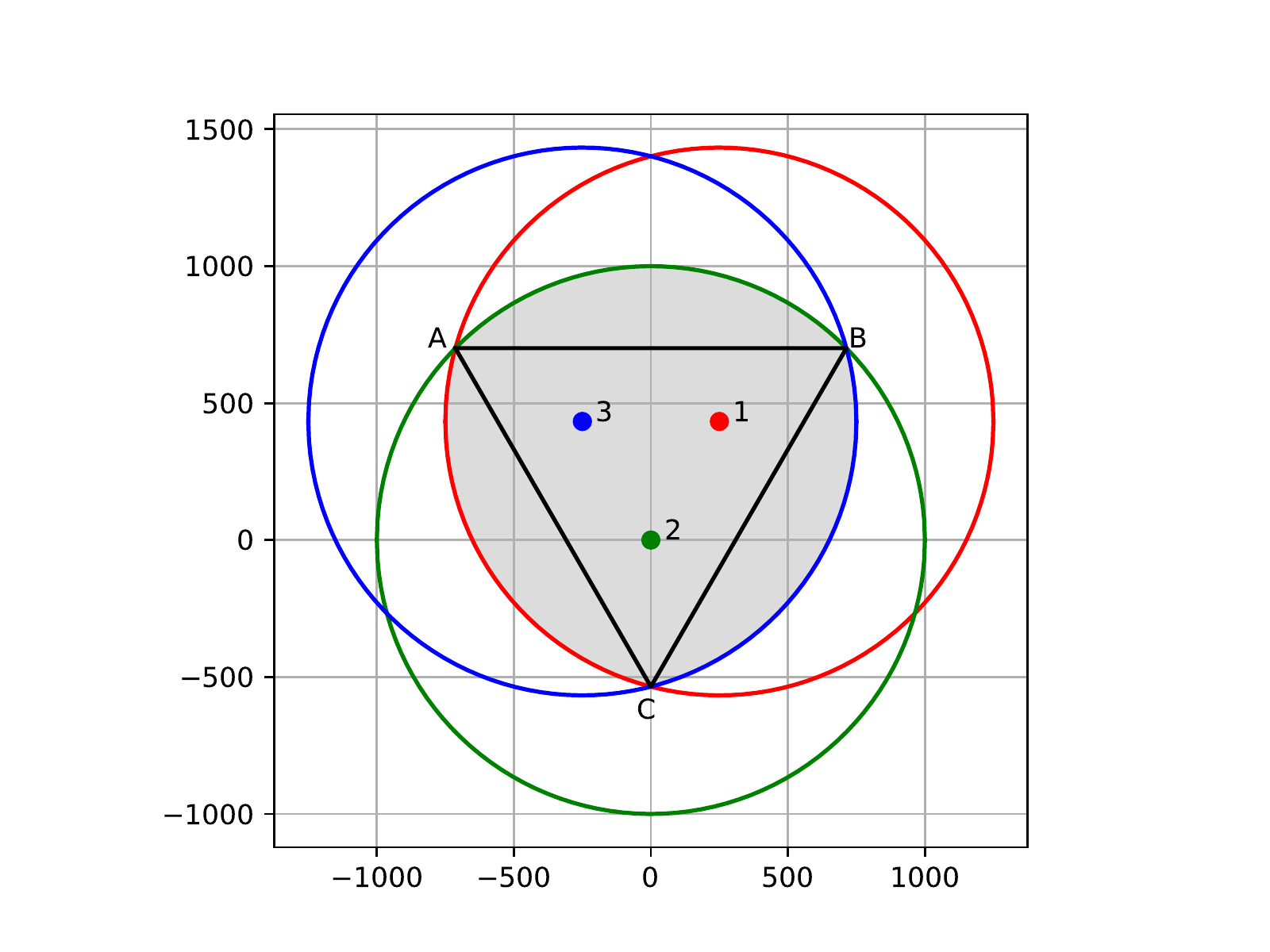}
	\captionsetup{width=.6\linewidth}
	\caption{The receivers' position are marked by numbers; the intersections between the contour of the receiving ranges are marked by letters. The shaded area is the usable receiving area $\mathcal{S}$. The distance between the 3 receivers is $0\leqslant x \leqslant R$.} 
	\label{fig:area}
\end{figure}

Let $R$ be the receiving radius of a receiver of all 3 receivers, and let $l$ be the range between the 3 receivers deployed in an equilateral setup. Without the loss of generality, let the position of one of the receivers be at $x_\mathrm{2}, y_\mathrm{2} = (0, 0)$. Consider the setup $x_\mathrm{1}, y_\mathrm{1} = (0.5l, 0.5\sqrt{3}l)$,  $x_\mathrm{3}, y_\mathrm{3} = (-0.5l, 0.5\sqrt{3}l)$. For $l=0$, the 3 receivers are placed in the same location and $\mathcal{S} = \pi R^2$. We derive the area of $\mathcal{S}$ for any $0\leqslant l \leqslant \sqrt{3}R$. Fig. \ref{fig:area} illustrates this setup. 
Referring to Fig.~\ref{fig:area}, the total usable area $\mathcal{S} = \mathcal{S}_\mathrm{T}+ 3\mathcal{S}_\mathrm{S}$, where $\mathcal{S}_\mathrm{T}$ denotes the area of triangle ABC and $\mathcal{S}_\mathrm{S}$ denotes the segments area between the cords AB, BC, CA. The intersection points of the two circles $(x - a_1)^2 + (y-b_1)^2 = R^2_1$ and $(x - a_2)^2 + (y-b_2)^2 = R^2_2$ are
\begin{eqnarray}
\begin{aligned}
x_\mathrm{1,2}=\frac{a_1+a_2}{2}+\frac{(a_\mathrm{2}-a_\mathrm{1})(R_\mathrm{1}^2-R_\mathrm{2}^2)}{2*l^2}\pm 2\frac{b_\mathrm{1}-b_\mathrm{2}}{l^2}d\\
y_\mathrm{1,2}=\frac{b_1+b_2}{2}+\frac{(b_\mathrm{2}-b_\mathrm{1})(R_\mathrm{1}^2-R_\mathrm{2}^2)}{2*l^2}\mp 2\frac{a_\mathrm{1}-a_\mathrm{2}}{l^2}d
\end{aligned}
\label{eq:intersect}
\end{eqnarray} 
where
\begin{eqnarray}
\begin{aligned}
d = 0.25\sqrt{(l+R_\mathrm{1}+R_\mathrm{2})(l+R_\mathrm{1}-R_\mathrm{2})(l-R_\mathrm{1}+R_\mathrm{2})(-l+R_\mathrm{1}+R_\mathrm{2})}\;.
\end{aligned}
\end{eqnarray}

For the case at hand, $R_\mathrm{1} = R_\mathrm{2} = R$ such that the distance between any two intersection points creating triangle ABC is $D = 0.5(\sqrt{3(4R^2-l^2)} -l)$. Thus,
\begin{eqnarray}
\begin{aligned}
&\mathcal{S}_\mathrm{T} = 0.25\sqrt3D^2\\
\mathcal{S}_\mathrm{S}= R^2\mathrm{sin^{-1}}&(\frac{D}{2R})-0.25D\sqrt{4R^2-D^2}\\
&\mathcal{S} =\mathcal{S}_\mathrm{T}+3\mathcal{S}_\mathrm{S}
\end{aligned}
\end{eqnarray} 

Fig. \ref{fig:toy} shows the coverage area for the distance between the receivers of $0\leqslant l \leqslant \sqrt{3}R$ for a number of GDOP values. 

\end{document}